\documentclass[manuscript]{acmart}
\AtBeginDocument{%
  }

\setcopyright{acmcopyright}
\copyrightyear{2023}
\acmYear{2023}
\acmDOI{XXXXXXX.XXXXXXX}

\acmPrice{15.00}
\acmISBN{978-1-4503-XXXX-X/18/06}

\usepackage{tabularx}
\usepackage{xcolor}
\usepackage{caption}
\usepackage{subcaption}
\usepackage{soul}
\usepackage{framed}
\usepackage{tablefootnote}

\begin{document}

%\title{Improving Clinical Imaging Interfaces with Cognition based Approaches}
%\title{Enhancing Clinical Imaging Interface by Utilizing Human Visual Processing Characteristics}
\title{Improving Clinical Imaging Systems using Cognition based Approaches}


\author{Kailas Dayanandan}
\affiliation{%
  \institution{Indian Institute of Technology, Delhi}
  %\streetaddress{1 Th{\o}rv{\"a}ld Circle}
  \city{Delhi}
  \country{India}}
\email{kailasd@gmail.com}

\author{Brejesh Lall}
\affiliation{%
  \institution{Indian Institute of Technology, Delhi}
  %\streetaddress{1 Th{\o}rv{\"a}ld Circle}
  \city{Delhi}
  \country{India}}
\email{brejesh@iitd.ac.in}

% FIXME 
%%\renewcommand{\shortauthors}{Trovato et al.}
\begin{abstract}
Clinical systems operate in safety-critical environments and are not intended to function autonomously; however, they are currently designed to replicate clinicians' diagnoses rather than assist them in the diagnostic process. To enable better supervision of system-generated diagnoses, we replicate radiologists' systematic approach used to analyze chest X-rays. This approach facilitates comprehensive analysis across all regions of clinical images and can reduce errors caused by inattentional blindness and under reading. Our work addresses a critical research gap by identifying difficult-to-diagnose diseases for clinicians using insights from human vision, enabling these systems to serve as an effective "second pair of eyes". These improvements make the clinical imaging systems more complementary and combine the strengths of human and machine vision. Additionally, we leverage effective receptive fields in deep learning models to present machine-generated diagnoses with sufficient context, making it easier for clinicians to evaluate them. %Our proposed approaches can potentially increase the adoption of deep learning systems, enhance clinician efficiency, and ultimately improve patient outcomes.
\end{abstract}

\begin{CCSXML}
<ccs2012>
   <concept>
       <concept_id>10003120.10003121.10003124.10010868</concept_id>
       <concept_desc>Human-centered computing~Web-based interaction</concept_desc>
       <concept_significance>500</concept_significance>
       </concept>
 </ccs2012>
\end{CCSXML}

\ccsdesc[500]{Human-centered computing~Web-based interaction}

\keywords{Generative AI, Healthcare, Chest X-Ray}

\received{14 September 2023}

\maketitle
 
\section{Introduction}

Asystems collaborating with clinicians have demonstrated improved performance and reduced diagnosis time \cite{leibig2022combining,calisto2022breastscreening,mbunge2023application}, emphasizing the need to enhance the adoption of AI-based systems. However, recent studies also indicate that human-AI collaboration can lead to lower performance, particularly when AI matches or surpasses human capabilities \cite{buccinca2021trust,green2019principles,park2019slow,fitzsimons2004reactance}. Human vision operates through a dual-process framework comprising an intuitive, fast-acting system (System 1) and a deliberate, analytical system (System 2). Failure to engage System 2's deliberate processing in clinical imaging tasks can result in diagnostic errors. For instance, research on inattentional blindness \cite{simons2010monkeying}, has been shown to affect clinical imaging \cite{hults2024inattentional,pinciroli2015unexpected,lum2005profiles,park2021eye,owattanapanich2017alk,viertel2012cervical,drew2013invisible}. Studies estimate the under readings to comprise 42\% of errors in clinical imaging \cite{kim2014fool}, including stopping after an initial instance is identified resulting in missing a second finding \cite{adamo2021satisfaction,berbaum2013satisfaction} (e.g., a patient with tuberculosis died from undiagnosed lymphoma \cite{owattanapanich2017alk}), and efforts to identify rare findings \cite{mitroff2014ultra,wolfe2007low}. These errors highlight clinicians' or radiologists' absence of deliberate and thorough System 2 analysis. A key objective of our paper is to make detailed analysis easier and efficient for doctors .

Human vision and machine vision differ significantly, with each having distinct biases and limitations. For instance, human vision often struggles to detect subtle patterns that deep learning models can easily identify. Radiologists view AI to be a "second pair of eyes" \cite{kohli2018cad} to enhance diagnostic accuracy. It is crucial to identify diseases that radiologists are prone to missing to improve overall diagnostic performance. This requires analyzing clinicians' cognitive processes to understand the sources of diagnostic errors for the deep learning models to focus on. There is a research gap in creating systems that identify and address hard-to-diagnose diseases to enhance the usability of AI systems in clinical settings. AI systems for clinical imaging often rely on web-based or conversational interfaces with limited display sizes \cite{yunxiang2023chatdoctor,xiong2023doctorglm,rajashekar2024human,fang2024experience,dayanandan2024enabling,gao2024taxonomy,kumar2023exploring}, which can prevent X-rays from being shown in full resolution. A wide range of diseases can be diagnosed from clinical images, making it challenging for human-computer interaction designers to understand the specific characteristics of each disease and determine the appropriate context around the affected region to be presented to clinicians for supervision.  In busy clinical settings, it is important to to have an intuitive interface that anticipates the information needs of clinicians  \cite{setchi2017exploring,dayanandan2024enabling} and provides the necessary context to improve usability and affective aspects of user experience, which is an open problem.

To make AI systems more complementary to clinicians, we developed a system that replicates the practices clinicians use to analyze X-rays, which have evolved to address human biases. Our approach enables clinicians to supervise and correct errors at each stage of their analysis and assists the clinicians in the various phases of the analytical process. We employ thematic analysis and inductive coding of clinicians' feedback from semi-structured interviews to identify common error patterns, and these diseases are prioritized while presenting the AI-generated diagnosis. Additionally, we provide diagnostic evidence and present affected regions at the appropriate resolution to facilitate easy verification. These enhancements reduce cognitive load for radiologists, strengthen the complementarity of AI-driven clinical imaging systems, and improve diagnostic accuracy by leveraging the capabilities of deep learning models. Our study seeks to improve clinician efficiency and diagnostic accuracy by developing a system that mirrors their analytical methods. Our major contributions include 

\textbf{(a)} Our approach assists clinicians in systematic analysis by replicating procedures designed to minimize human biases, which was preferred by clinicians for computer-based diagnosis. Instead of directly analyzing clinical images to generate findings, our method allows clinicians to supervise the diagnostic process using established procedures. This methodology can enhance clinicians' diagnostic accuracy and operational efficiency. 

\textbf{(b)} Our study observes that inattentional blindness, inherent limitations of human vision, and operating environments can contribute to diagnostic errors made by clinicians. Deep learning models are better at detecting small affected areas and subtle patterns often overlooked by the human eye, making these systems complementary to radiologists. 

\textbf{(c)} Our study leverages the concept of effective receptive fields in deep learning models to automatically determine the context around the affected area required to evaluate machine diagnosis for different diseases. This can help avoid the searching, scrolling, and zooming required to get the context into view for diagnosing a disease on web-based systems.

\section{Related Work}

\begin{figure*}[ht]
    \centering
    \subfloat[ \label{fig:navon-dataset}]{\fbox{\includegraphics[height=1.45cm]{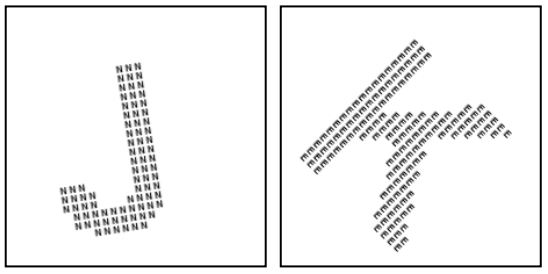}}} \hfill
    \subfloat[ \label{fig:navon-consistency}]{\fbox{\includegraphics[height=1.45cm]{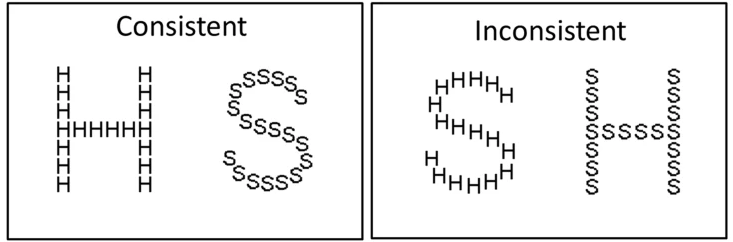}}} \hfill
    \subfloat[ \label{fig:image-abstraction}]{\fbox{\includegraphics[height=1.45cm]{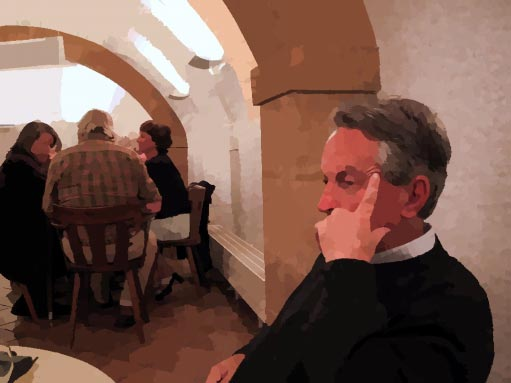}}} \hfill
    \subfloat[ \label{fig:stylized-imagenet-example}]{\fbox{\includegraphics[height=1.45cm]{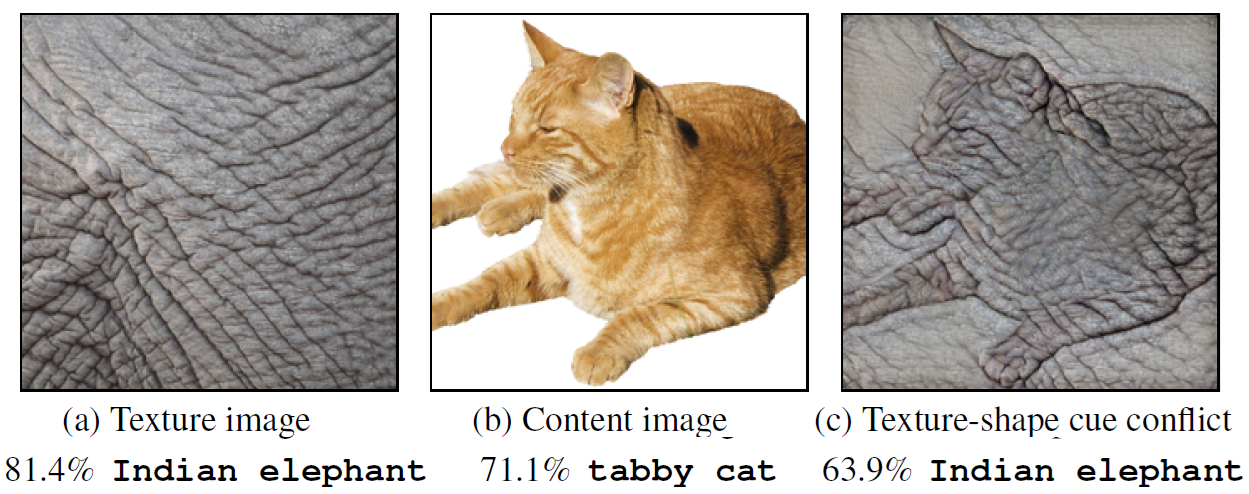}}} \hfill 
    
    \subfloat[ \label{fig:dual-gen-a}]{\fbox{\includegraphics[height=1.75cm]{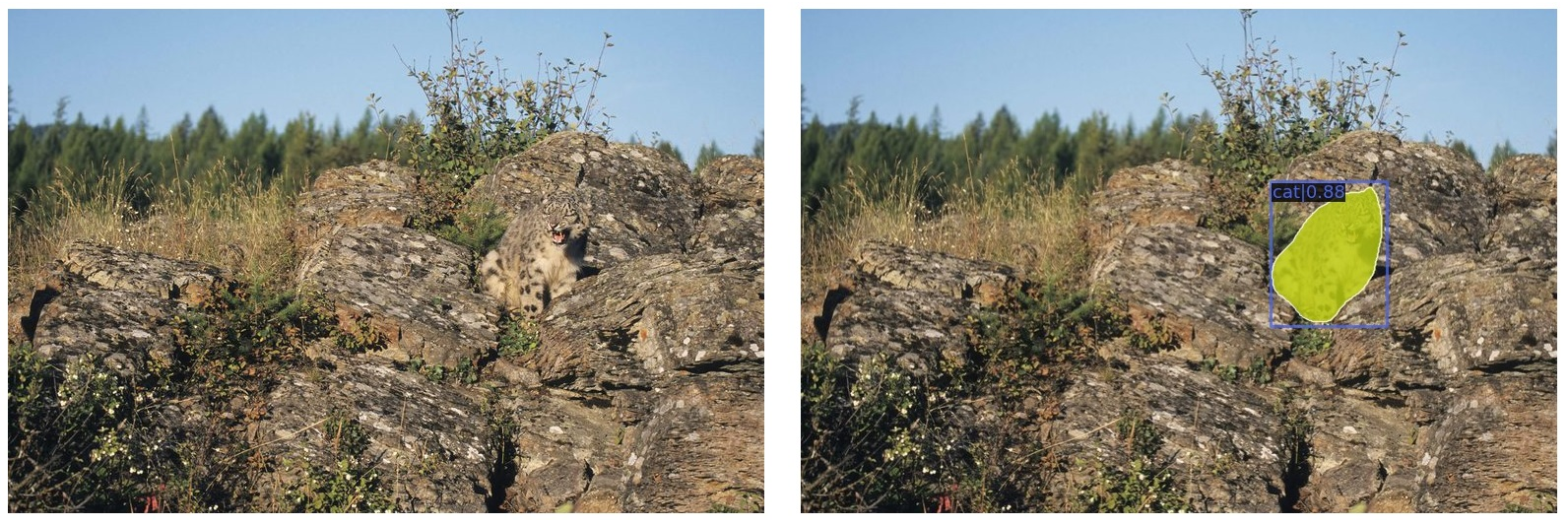}}} \hfill    
    \subfloat[ \label{fig:gen-a}]{\fbox{\includegraphics[height=1.75cm]{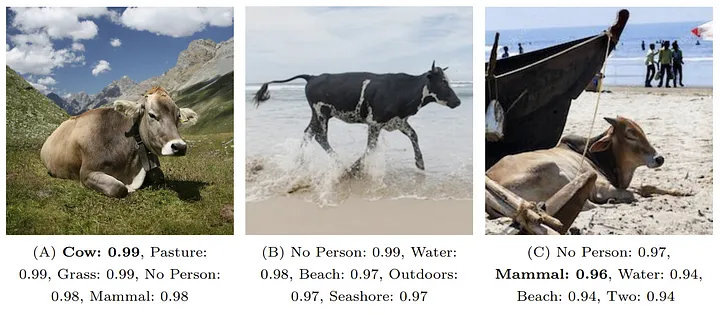}}} \hfill        
    \subfloat[ \label{fig:imagenet-a}]{\fbox{\includegraphics[height=1.75cm]{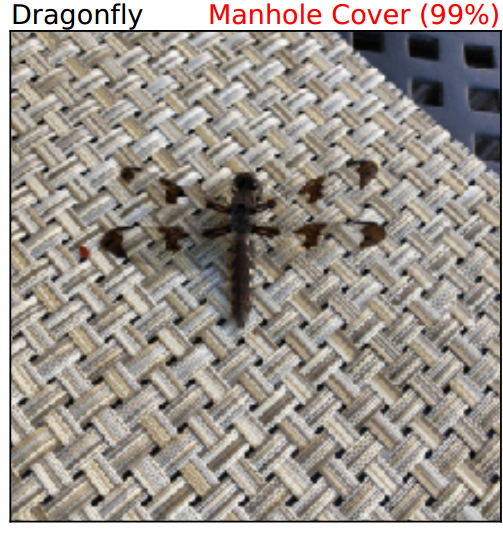}}} \hfill        
    \subfloat[ \label{fig:stylized-adversarial}]{\fbox{\includegraphics[height=1.75cm]{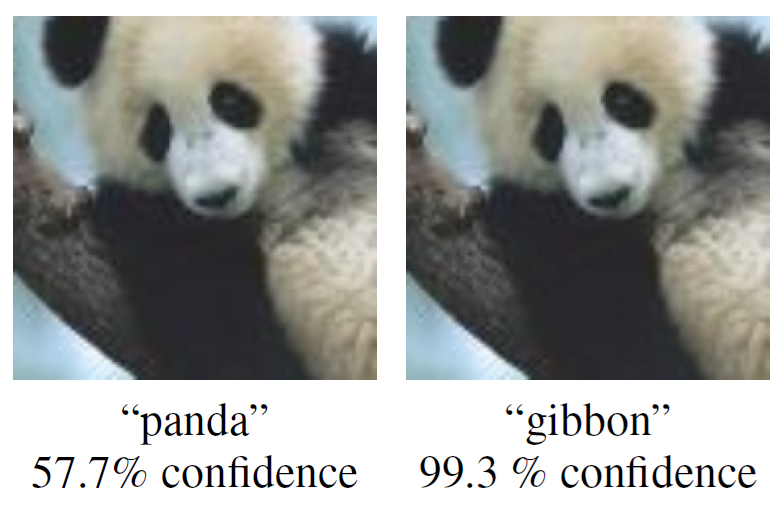}}} \hfill     
    \caption{(a) Navon Dataset has images with a large letter rendered in small copies of some other letter \cite{navon1977forest} and adapted with rotation at angles between -45 and 45 degrees \cite{hermann2020origins} (b) Considerable differences exist based on formation \protect\cite{gerlach2018navon} (c) An example from Human Confusion Dataset explaining the dual thinking framework where the cover on table is wrongly inferred first as a cat   \protect\cite{dayanandan2024dual}. Dual thinking framework is being studied in perceptual \cite{dayanandan2024dual} and electro-physiological studies \cite{vanrullen2007power,grootswagers2019representational,kreiman2020beyond,thorpe1996speed,van2020going,tang2018recurrent} (d) Human vision does not require complete information and the image shows this perceptual abstraction \cite{decarlo2002stylization} (e) Human vision depends on shape whereas deep learning models rely on texture shown in an example from Stylized ImageNet Dataset (SIN) \cite{geirhos2018imagenet}  (f) Focus on texture can help in identifying objects that are difficult for human vision using an example from Human Confusion Dataset \protect\cite{dayanandan2024dual} (g) Deep learning models also takes into account other features that can improve the accuracy which though affects generalization can be helpful in medical imaging to use more features \protect\cite{beery2018recognition} (h) The deep learning models are prone to errors that human vision are not prone to (an example from ImageNet-A) \protect\cite{hendrycks2021natural} (i) Deep learning models are prone to adversarial attacks which are not likely to be present in safety critical clinical setting \protect\cite{goodfellow2014explaining}.}    
    \label{fig:VisualTheories}
\end{figure*}

Clinical systems are becoming increasingly surpassing human performance in various tasks. However, studies have shown that the performance of human-AI collaboration often does not improve as expected. This has led to a focus on "algorithm-in-loop" approaches \cite{green2019principles}, where humans incorporate algorithmic inputs while making final decisions. Cognitive forcing techniques have evolved to prevent over-reliance on AI systems for routine tasks \cite{buccinca2021trust}. These methods include having AI and user make independent decisions and comparing them \cite{green2019principles}, slowing down the presentation of AI predictions to allow time for reflection \cite{park2019slow}, and allowing users to choose when to view recommendations \cite{fitzsimons2004reactance}. However, having clinicians independently analyze and verify AI-generated diagnoses or delaying the presentation of AI findings reduces the efficiency gains that AI could offer. The lack of significant improvements in efficiency makes it challenging to justify the investment in such systems, especially considering the already high accuracy of radiologists. An approach that complements the radiologist and collaborates to perform System 2 analysis that can ensure a comprehensive evaluation of all aspects of an image is still an open problem.

Human vision is robust and generalizable, relying on a coarse-to-fine processing approach \cite{vanrullen2007power,van2020going,mohsenzadeh2018ultra,daniel2017thinking,chen2021cooperative}, whereas machine vision focuses on texture and can be brittle, often susceptible to adversarial attacks \cite{muttenthaler2024improving,muttenthaler2024improving,geirhos2020shortcut,dapello2020simulating,linsleylearning,linsley2020stable,fel2022harmonizing}. Clinical images, being synthetic, display patterns distinct from natural images, which human vision has not evolved to interpret effectively. While research consistently demonstrates that human-AI collaboration outperforms human performance alone, it often remains less effective than AI functioning independently \cite{bansal2021does,green2019principles,oviatt2007multimodal}. For example, a study by Michelle et al. indicates that collaboration is unproductive when humans and algorithms exhibit similar decision-making patterns \cite{vaccaro2019effects}. Recent studies highlight the need to understand human biases in decision-making and develop systems that effectively complement human capabilities \cite{rastogi2023investigating}. A nuanced understanding of disease characteristics, combined with insights into human biases (Fig. \ref{fig:VisualTheories}) can help identify diseases that clinicians might overlook \cite{kohli2018cad}. Deep learning models can uncover novel features that enhance their ability to complement clinicians, thereby improving overall performance through collaboration. Our study addresses a research gap by identifying common sources of errors and their connection to cognitive processes in human vision, allowing these errors to be effectively prioritized when presented to the user.

Clinical imaging systems align with the "high control, high automation" category outlined in Shneiderman's human-centered AI framework \cite{shneiderman2022human}. Therefore, enhancing user interaction and improving diagnostic accuracy are crucial to achieving responsible automation in these safety-critical systems. The analysis of clinical images involves navigating a complex, sequential decision-making process compared to simpler tasks analyzed in current studies \cite{buccinca2021trust}. For instance, in chest X-rays, clinicians initially assess factors such as rotation, inspiration, and exposure before systematically examining regions related to the airway, breathing, circulation, diaphragm, and other regions. Systems that provide evidence for each step in the diagnostic process can better support clinicians' deliberate systematic (System 2) analysis. These systems should aim to help clinicians work more effectively with AI \cite{thimbleby2021fix}, whereas current clinical systems provide the final diagnosis, effectively replacing radiologists. They should be designed to enhance supervision, prevent over-reliance on AI, and ensure that detailed analysis is not overlooked \cite{lai2019human}. Research indicates that clinicians favor decision support systems that mirror their analytical workflows for patient care decisions \cite{yang2023harnessing}. Incorporating advancements in large language models with existing accurate diagnostic methods effectively for easier supervision and improving accuracy is still an open problem.

Deep learning models have been employed as computational models of human vision \cite{vogelsang2018potential,kreiman2020beyond,saxe2021if,kim2019disentangling,lake2017building}, particularly in studies that explore the significance of receptive fields \cite{vogelsang2018potential,yan2020revealing,kubilius2016deep}. In web-based interfaces, including conversational interfaces, limited screen space requires clinicians to manage the displayed information effectively \cite{dayanandan2024enabling}. There is a research gap in understanding the contextual information needed for diagnosing diseases based on their characteristics or criteria for diagnosis. Determining the amount of contextual information needed for disease diagnosis automatically using the model's effective receptive field is still an open problem.

\section{Methods}

\begin{figure}[h]
\begin{center}   
    \includegraphics[width=\textwidth]{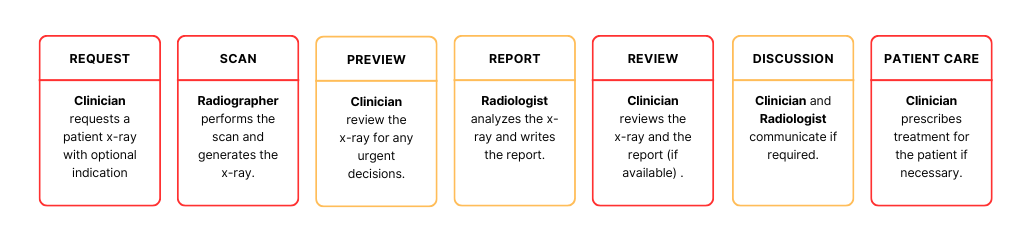}   
\end{center}
	\caption{Radiology workflow from the clinician to final patient care. In our user trials, we observe that the preview, report, and discussion do not happen in typical rural settings. The steps in red are essential steps, while those with orange border may not be present.}
    \Description{Radiology workflows.}
\label{fig:mm-proposed}
\end{figure}

Radiology workflows starts with a clinician requesting a chest X-ray with an optional indication to the radiologist. The radiographer captures the X-ray, and the radiologist examines the X-ray along with the indication provided by the clinician and writes the report. If needed, the clinician reviews this report and discusses it with the radiologist before recommending treatment \cite{yildirim2024multimodal} (Fig.\ref{fig:mm-proposed}). Our primary focus is to design the workflow better to improve the usability of interfaces by making it easier to analyze the clinical images, which brings to our research questions for our study

\vspace*{0.5cm}
\begin{quote}
\textit{
RQ. 1 : How to make the diagnostic systems complementary for radiologists ?
}
\end{quote}
%\vspace*{0.5cm}
\begin{quote}
\textit{
RQ. 2 : How do we identify diseases that clinicians are likely to miss out?
}
\end{quote}
%\vspace*{0.5cm}
\begin{quote}
\textit{
RQ. 3 : How to make systems easier for clinicians to supervise ?
}
\end{quote}
%\vspace*{0.5cm}

We designed a system as per our proposed approach to replicate the radiologist’s diagnostic process and present machine-generated diagnoses in a way that reduces cognitive load and enhances efficiency. This design enables clinicians to focus on supervising the machine’s output rather than conducting a complete diagnosis themselves. To evaluate the effectiveness of this approach, we conducted a qualitative study with clinicians. We engaged in discussions to identify the characteristics of diseases that radiologists are more likely to overlook. Additionally, we leverage deep learning models to determine the contextual information necessary for clinicians to assess the diagnoses.

\subsection{Study Design}
In our qualitative study, we conducted semi-structured interviews with sixteen clinicians to evaluate the effectiveness of our approach. For our method to be impactful, the analysis of imaging modalities must be a complex, sequential process that is widely used by clinicians and radiologists. The interviews were structured around our research questions and covered the following topics: (a) the clinicians' process of analyzing X-rays, enabling us to observe the nuances of their approach and ask follow-up questions to gain deeper insights when necessary, (b) the use of standard ABCDE method for analysis in their clinical practice, (c) the usefulness of our proposed diagnostic system, (d) general questions about diseases they find challenging to identify, (e) the characteristics and common patterns in diseases that make them difficult to diagnose, (f) the usefulness of interface could help identify diseases that are likely to be missed out, and (g) the context near the affected region that clinicians need for assessing diseases. The interviews concluded with a discussion of the tool's usefulness in clinical settings. The participants were recruited either from the institute hospital or were referred by the lab members to which the authors belong. Ten participants had an advanced degree, and the remaining doctors had bachelor's degrees. Similarly, eight doctors had over 10 years of experience. The specialization, qualification, and years of experience of participants are shown in Table \ref{tab:participant}.

\begin{table}[h]
  \caption{Participant Details}
  \label{tab:participant}
  \centering
  %\begin{tabularx}{0.85\linewidth}{p{15pt} p{120pt} p{28pt} p{25pt}}
  \begin{tabular}{llll}
    \toprule
    ID & Specialization \& Qualification & Experience & Location \\
    \midrule
    \textbf{P1} & Cardiac Surgeon (MS, MCh) &  10+ & Rural \\  
    \textbf{P2} & General Surgeon (MS) &  24  & Urban \\  
    \textbf{P3} & General Physician (MBBS) &  35 & Urban \\ 
    \textbf{P4} & Anesthesiologist (DNB) &  10+  & Urban \\
    \textbf{P5} & Emergency Physician (DEM) &  18+  & Urban \\   
    \textbf{P6} & Emergency Physician (DEM, DGM) &  15+  & Urban \\      
    \textbf{P7} & Psychiatrist (MD, DNB) &  4+  & Urban \\        
    \textbf{P8} & Family Physician (MD, DNB) &  15+  & Urban \\    
    \textbf{P9} & General Physician (MBBS) &  3+ & Urban \\     
    \textbf{P10} & General Physician (MBBS) &  3+ & Urban \\     
    \textbf{P11} & Oncologist (MS) &  3+  & Rural \\  
    \textbf{P12} & Pediatrician (MD) &  10+  & Urban \\
    \textbf{P13} & General Physician (MBBS) &  5+  & Urban \\
    \textbf{P14} & General Physician (MD) &  4 & Rural \\  
    \textbf{P15} & General Physician (MBBS) &  4  & Rural \\    
    \textbf{P16} & General Physician (MBBS) &  3+ & Rural \\    
  \bottomrule
\end{tabular}
\end{table}

\subsection{Data Collection and Analysis} 
\begin{figure*}[ht]
    \centering

    \subfloat[ \label{fig:original1-abcde}]{\includegraphics[width = 2.0cm]{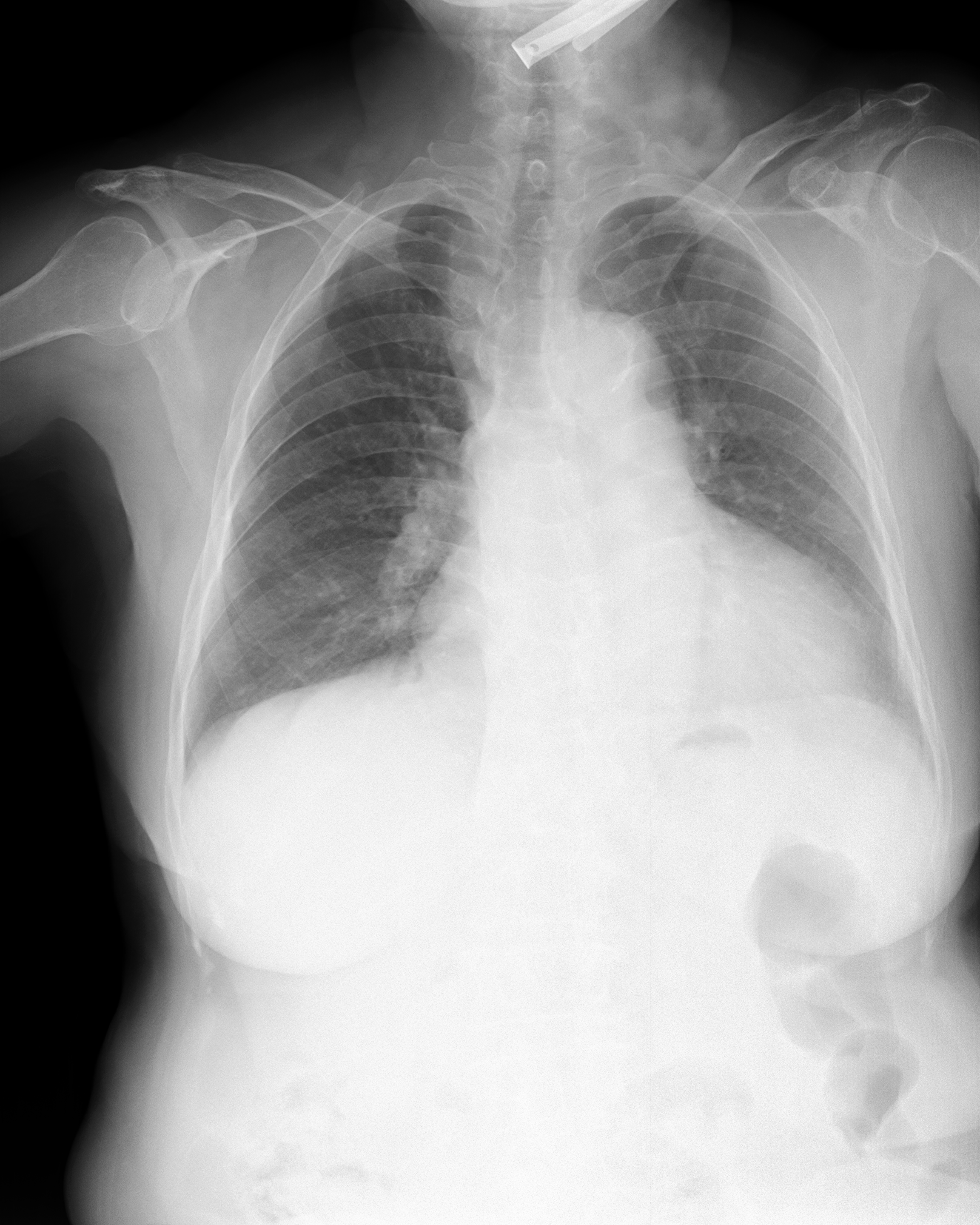}} \hfill
    \subfloat[ \label{fig:a1-abcde}]{\includegraphics[width = 2.0cm]{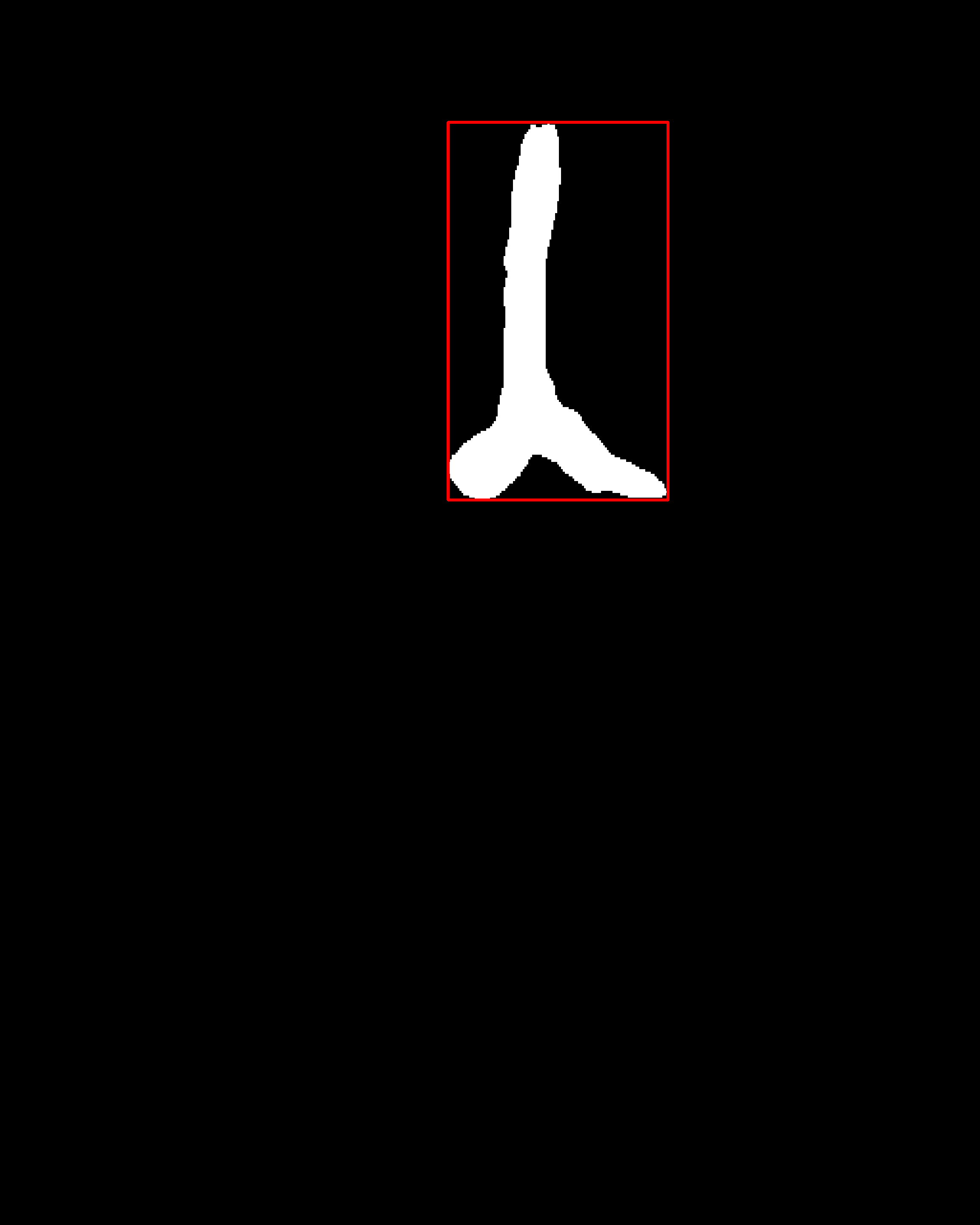}} \hfill
    \subfloat[ \label{fig:b1-abcde}]{\includegraphics[width = 2.0cm]{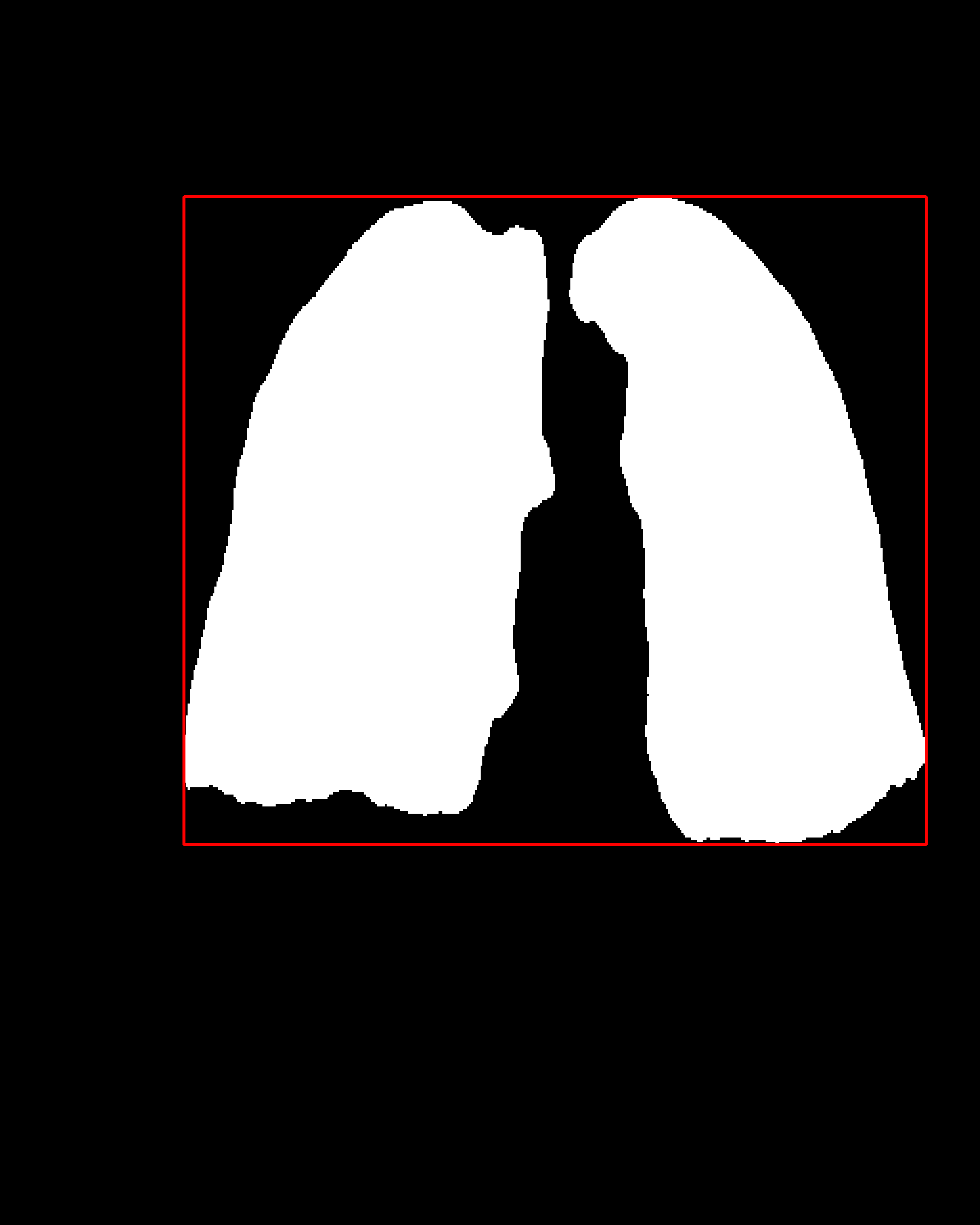}} \hfill
    \subfloat[ \label{fig:al1-abcde}]{\includegraphics[width = 2.0cm]{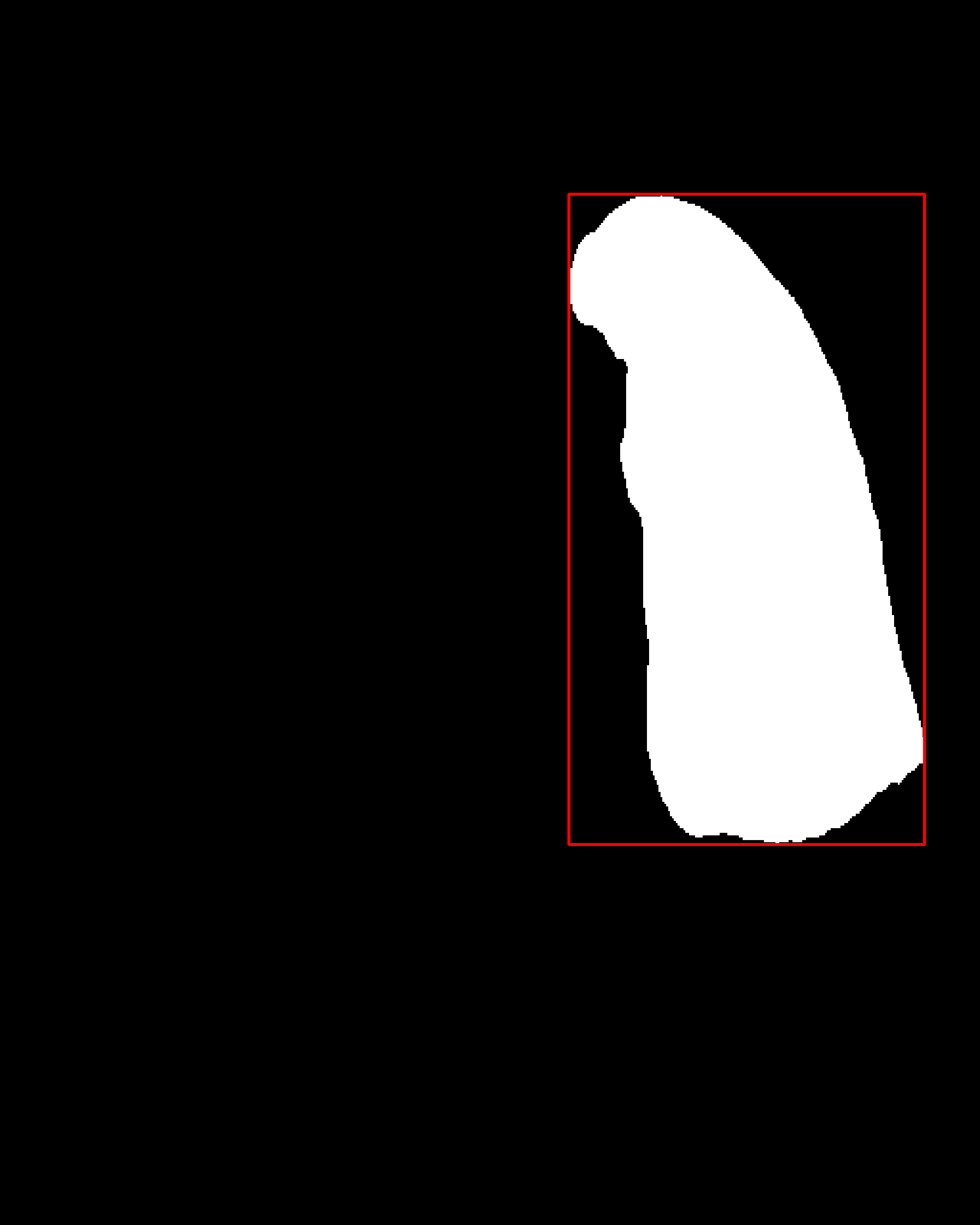}} \hfill
    \subfloat[ \label{fig:ar1-abcde}]{\includegraphics[width = 2.0cm]{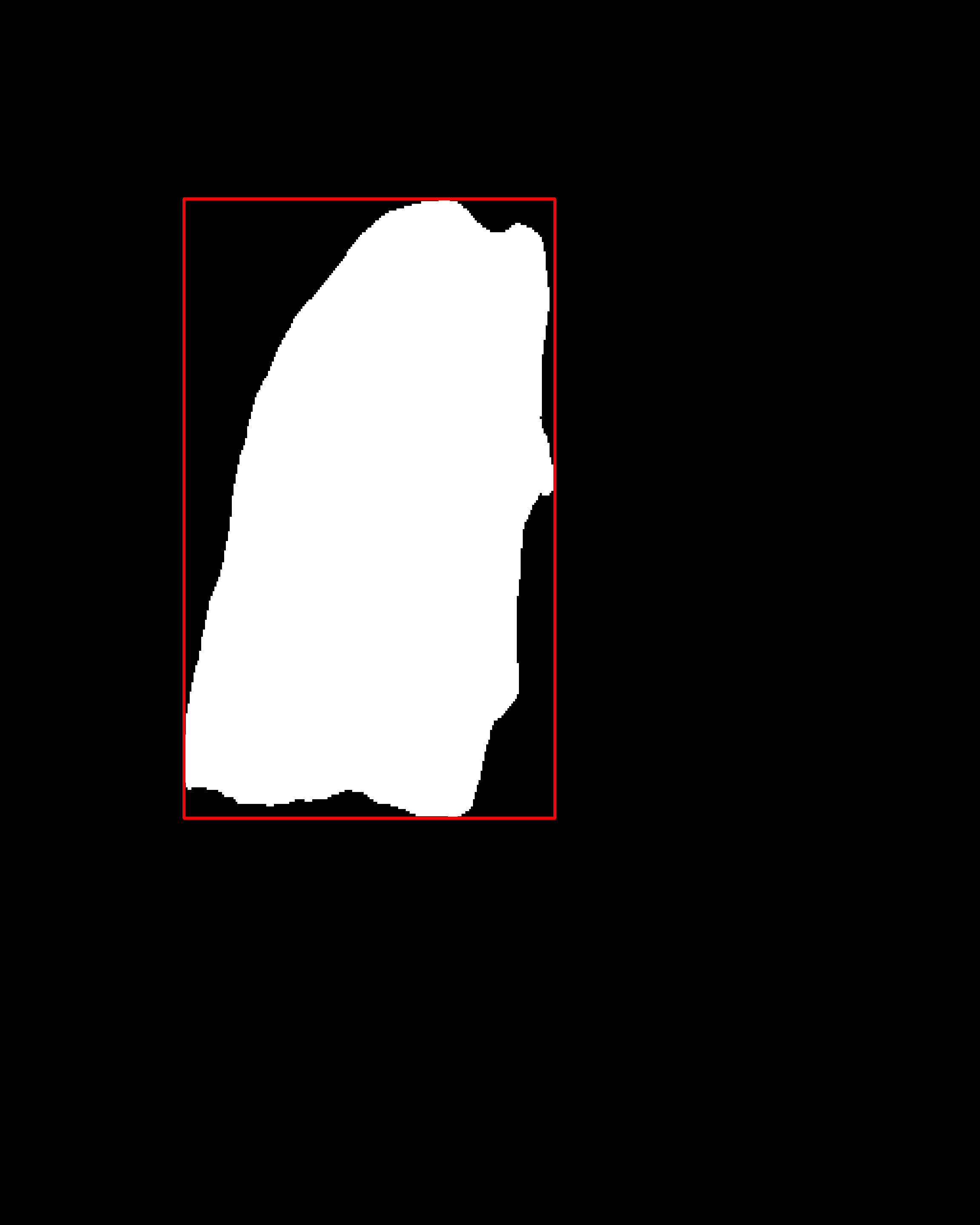}} \hfill    
    \subfloat[ \label{fig:c1-abcde}]{\includegraphics[width = 2.0cm]{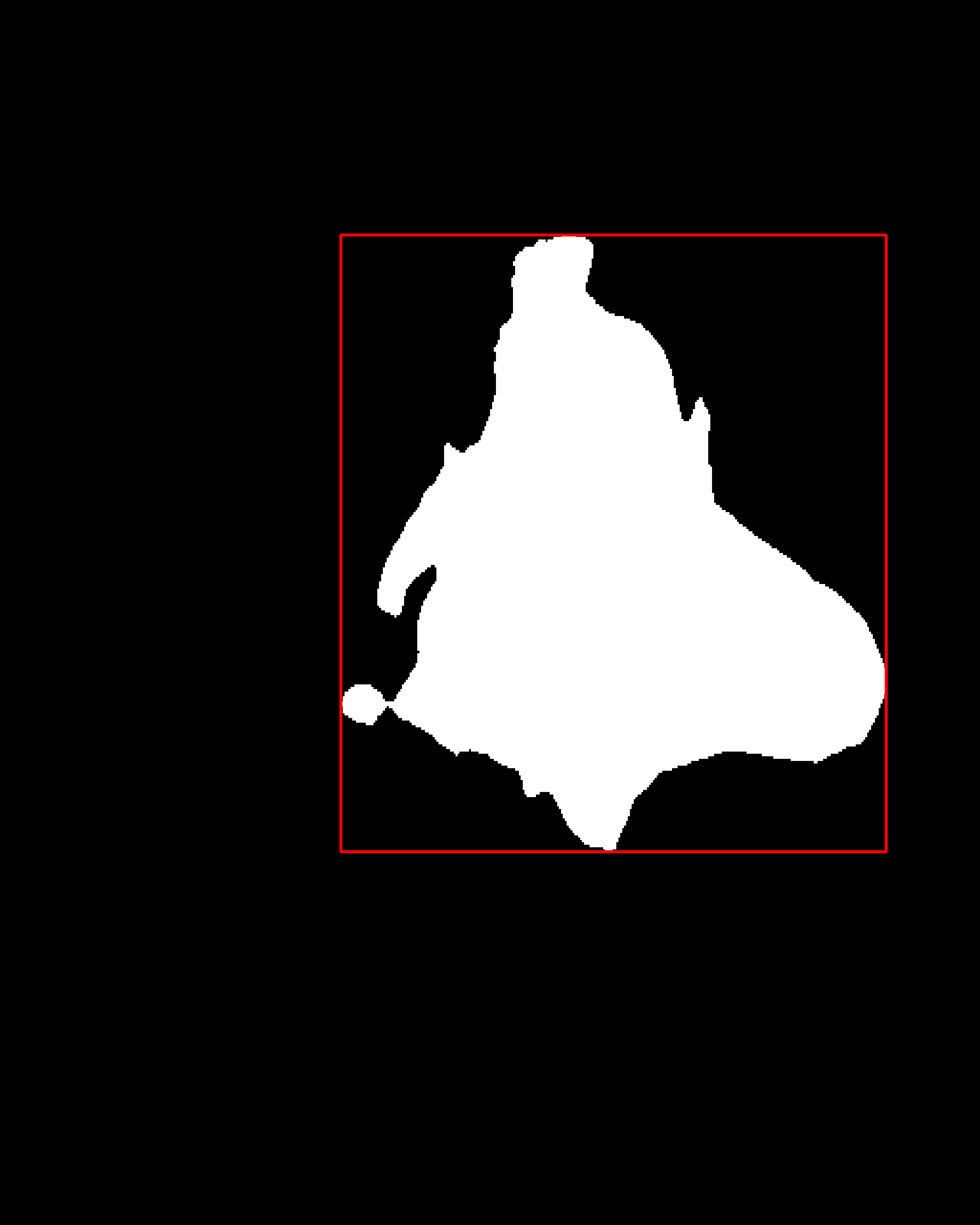}} \hfill
    \subfloat[ \label{fig:d1-abcde}]{\includegraphics[width = 2.0cm]{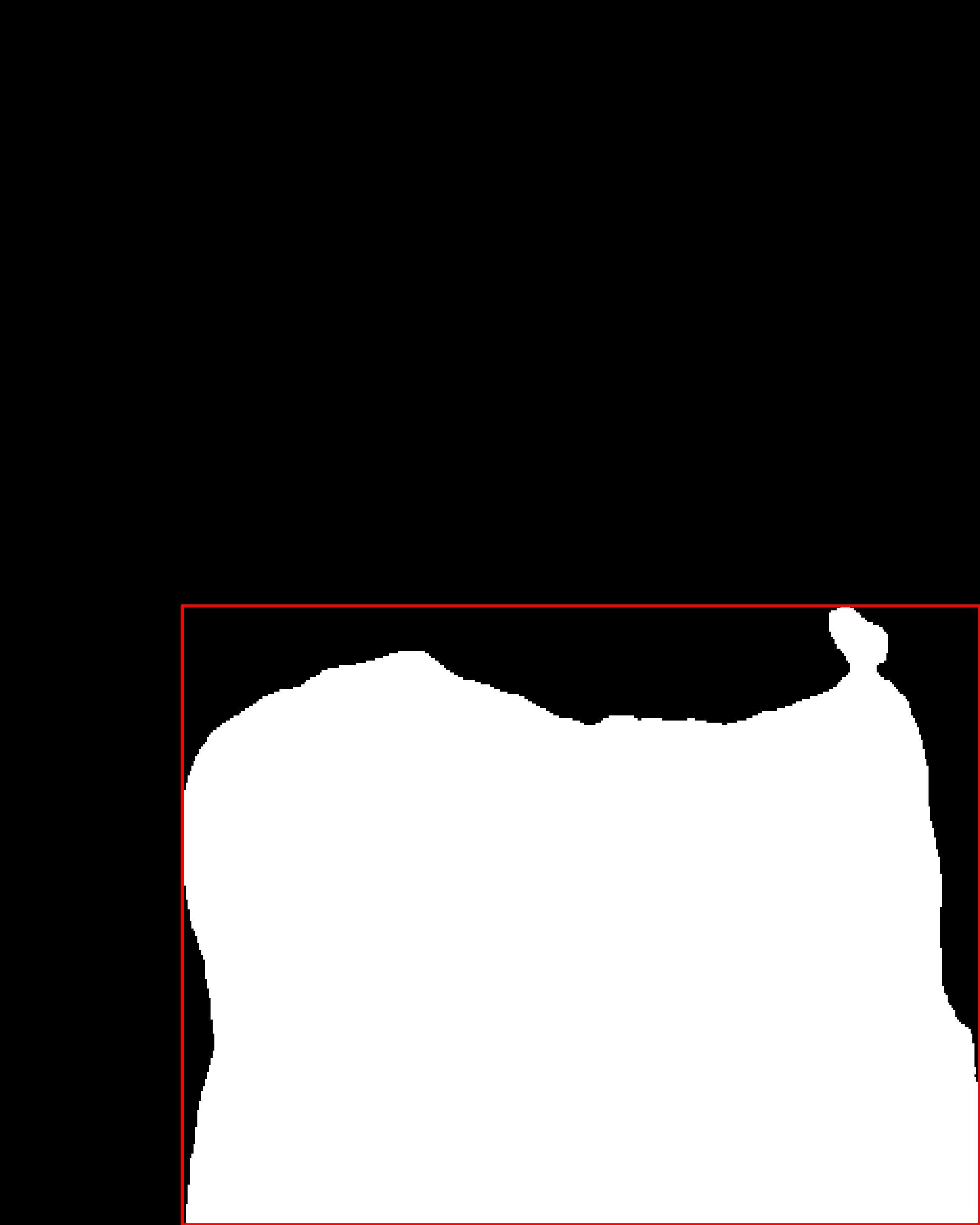}} \hfill
    
    \caption{First row contains the original x-ray and various regions generated and shown on the x-ray, while second row shows the masks generated for different regions in ABCDE approach. We use a random x-ray from VinDr-CXR dataset (a) Original image (b) \textbf{A}irway consisting of trachea and mediastinal width (e) \textbf{B}reathing showing lung fields. This also identifies relevant region to be show for evaluating cardiomegaly in Fig.\ref{fig:abcde_process} (c) Left Lobe (d) Right Lobe (f) \textbf{C}irculation showing the heart and related regions (g) \textbf{D}iaphragm}    
    \label{fig:dl-abcde}
\end{figure*}

We started with clinicians demonstrating their process of analyzing an X-ray, which also captures the elements an ethnomethodological study can capture. We then presented the clinicians with mockups to elicit their feedback. We conducted a thematic analysis of their responses to identify common patterns in diseases that are challenging to diagnose and the reason for preferring our tool. The interviews lasted between 30 minutes to 1 hour. %Additionally, we use three extrinsic datasets, VQA-RAD - a question-answering dataset containing questions asked by clinicians on radiology images and their reference answers \cite{lau2018dataset} to understand clinicians' requirements and VinDr-CXR, MIMIC-CXR \cite{johnson2019mimic} datasets that contain chest x-ray's and their disease diagnosis labels. 
We implemented the prototype's user interface in HTML (Fig.\ref{fig:abcde_process}). We extracted the different regions with deep-learning models for the Airway, Breathing, Circulation, and Diaphragm to ensure feasibility (Fig.\ref{fig:dl-abcde}). We used sample X-rays, corresponding diagnosis, and ground truth annotations from the VinDr-CXR dataset to show diseases and associated affected regions (Fig.\ref{fig:zoom-eval}).

\subsection{Extrinsic Datasets} 
\label{extrinsic-appendix}
In our study, we use three extrinsic datasets with different details for x-rays. We use the question answering dataset (VQA-RAD) to check whether the information provided in the workflow is relevant and to understand the changes in academic and clinical settings. MIMIC-CXR dataset contains associated reports and disease labels and this dataset is used in experiments for finding the context required for presenting to user.

\subsubsection{Question Answer Dataset } 
VQA-RAD is a publicly available visual question-answering dataset containing 2248 questions asked by clinicians about radiology images and their reference answers \cite{lau2018dataset} on 104 head axial single-slice CTs or MRIs, 107 chest x-rays, and 104 abdominal axial CTs. 

\subsubsection{Medical Report Dataset } 
MIMIC-CXR contains 3,77,110 chest x-rays and associated semi-structured free-text radiology reports from 227,835 imaging studies conducted on 65,379 patients at the Beth Israel Deaconess Medical Center Emergency Department in Boston \cite{johnson2019mimic} and has many derivatives with additional information \cite{tam2020weakly,lanfredi2021reflacx,boecking2022making,liao2021pulmonary}.

\subsubsection{Disease Annotated Dataset } 
VinDr-CXR is a publicly available dataset with 18,000 postero-anterior (PA) chest X-rays collected from Hospital 108 and the Hanoi Medical University Hospital \cite{nguyen2022vindr}.

\section{Findings}

In the first part of this section, we focus on the effectiveness of the proposed think-along workflow; the second part explores diagnostic errors made by clinicians and their connection to cognitive limitations; and the final part discusses improved presentation of information that can help evaluate the machine diagnosis easily.

\subsection{Image Analysis (RQ:1)}

In order for the proposed method to be beneficial, the standardized analysis approach must be widely adopted by clinicians, and the interviews provided insights into their method of analysis. This section explores clinicians' process to analyze chest X-rays, which involves a complex, sequential examination of different regions. Additionally, we discuss our approach, which integrates a standardized analysis process using AI to support clinicians in their deliberate (system 2) analysis.

\textit{\textbf{(1)}} Rotation, Inspiration, Projection and Exposure (RIPE) is initially used to assess the image quality. The questions posed by the students to a potential system, as captured in the VQA-RAD dataset cover all of these initial evaluation criteria \textit{"was the patient positioned appropriately without tilting? (Q:41)"}, \textit{"Was this chest x ray taken in PA format? (Q:95)} \textit{"Does this represent adequate inspiratory effort? (Q:12)"}, \textit{"Are there at least 8 ribs visible for good inspiratory effort? (Q:57)"} and \textit{"Was this taken in PA position? (Q70)}. This information is not included in the final report but serves as an intermediate step, indicating that students use the system to support their analytical process in reaching a final diagnosis. Clinicians engage with the system as a tool that assists them through different stages of analysis.

\textit{\textbf{(2)}} The participants mentioned the identification of the view and the position of the left and right lungs as important \textit{".. the right lung and left lung should be correctly identified.. it is also important in medico-legally and clinically both.. (P11)"}. However the method of identification is not consistent across the participants, for example many senior participants relied on cardiac shadows \textit{".. we do no go by the left right marking.. we go by the gastric shadows.. and cardiac shadows .. unless the patient has a dextro rotation .. (P6)"}, while some relied on ribs, clavicle or the markings \textit{".. the view and the left and right will mostly be marked on the x-ray.. (P10)"}, hence  it may not be possible to fully replicate this step, but the system can indicate the identified view.

\textit{\textbf{(3)}} The analysis of chest x-ray is recommended to be conducted sequentially for different regions Airways, Breathing, Circulation, Diaphragm and Extras (ABCDE method). The participants mentioned that the sequential analysis by focusing on these regions to be helpful \textit{".. we have to follow thoroughly.. else I might miss something.. I should have a broad spectrum of diagnosis in my mind.. I would prefer looking at the x-ray directly.. if it is on the computer.. I would prefer this.. (P6)"}, and \textit{".. this process is fine.. (P2)"}. The participants also mentioned that the analysis can be more accurate with one participant making a comment while observing the expanded view of airway (Fig.\ref{fig:abcde_process}) \textit{" .. you can observe them better.. it could pick up small things that doctors would miss .. (P2)"}.

\textit{\textbf{(4)}} Many participating clinicians mentioned that they may not always receive an accompanying report with the chest X-ray \textit{".. most of the time you do not get a report with the x-rays.. ..it is different from what happens in cities.. in small villages like where I am.. we do not get the reports.." (P15)} and some of the participants choose to do the analysis themselves \textit{".. I would want to go through from top to bottom.. I would want to do it on my own .. (P15)"}. As noted in previous studies \cite{kohli2018cad}, there is potential for AI to serve as a second pair of eyes by identifying diseases that clinicians are likely to overlook \textit{".. I got my CT scan done very recently.. some of findings were not there.. ENT doctor .. could figure out what the radiologists missed.. there can be good people and bad people.. it is a very subjective thing.. (P15)"}. The clinicians also tend to focus their analysis to potential areas as per the symptoms \textit{" .. we do not analyze using ABCDE approach .. we look at the x-ray with the symptoms the patient in mind.. .. we do not have time to do look at the x-ray exhaustively.. (P10)"}, however, the participants were interested in systems that can detect any missed diseases \textit{".. if the app can find these nodules.. .. and categorize if it is infective or malignant.. then it is helpful.. (P11)"}.

\begin{figure*}[h]
\begin{center}   
    \subfloat[ \label{fig:ABCDE-original}]{\includegraphics[height=2.5cm]{images/ABCDE/051132a778e61a86eb147c7c6f564dfe.png}} \hfill %\qquad  %\hfill
    \subfloat[ \label{fig:ABCDE-together}]{\includegraphics[height=2.5cm]{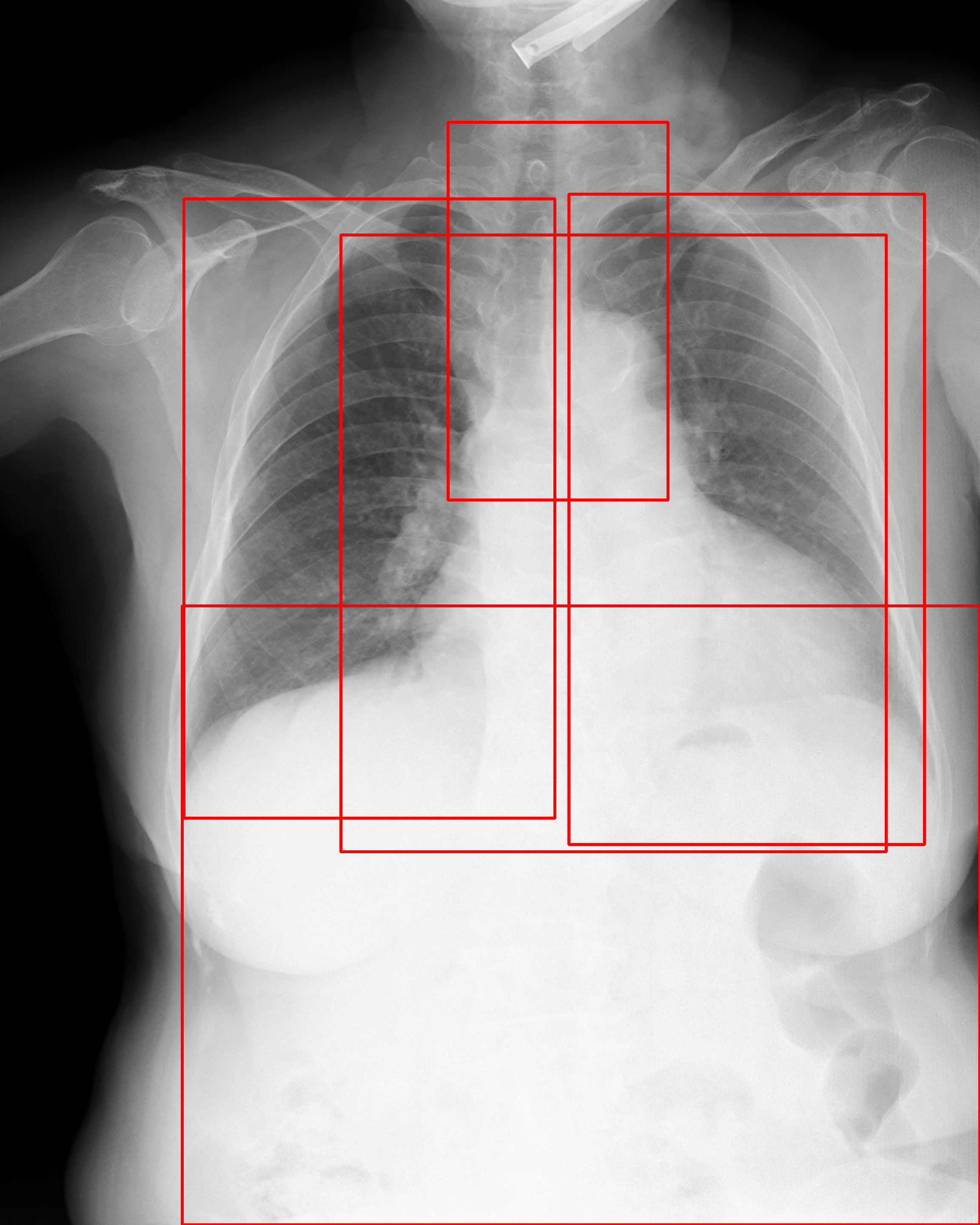}} \hfill %\qquad  %\hfill
    \subfloat[ \label{fig:ABCDE-original}]{\includegraphics[height=2.5cm]{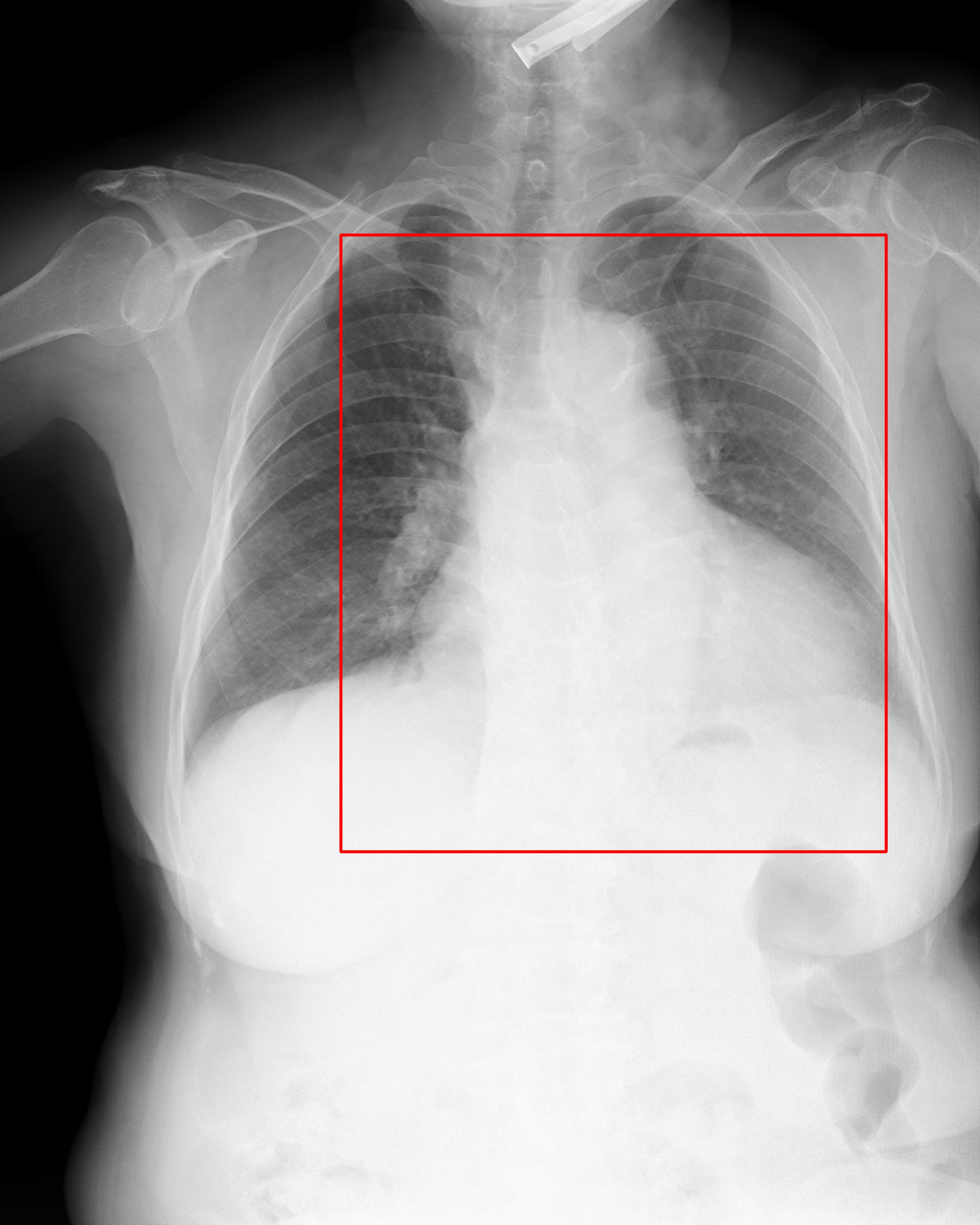}} \hfill %\qquad  %\hfill
    \subfloat[ \label{fig:ABCDE-original}]{\includegraphics[height=2.5cm]{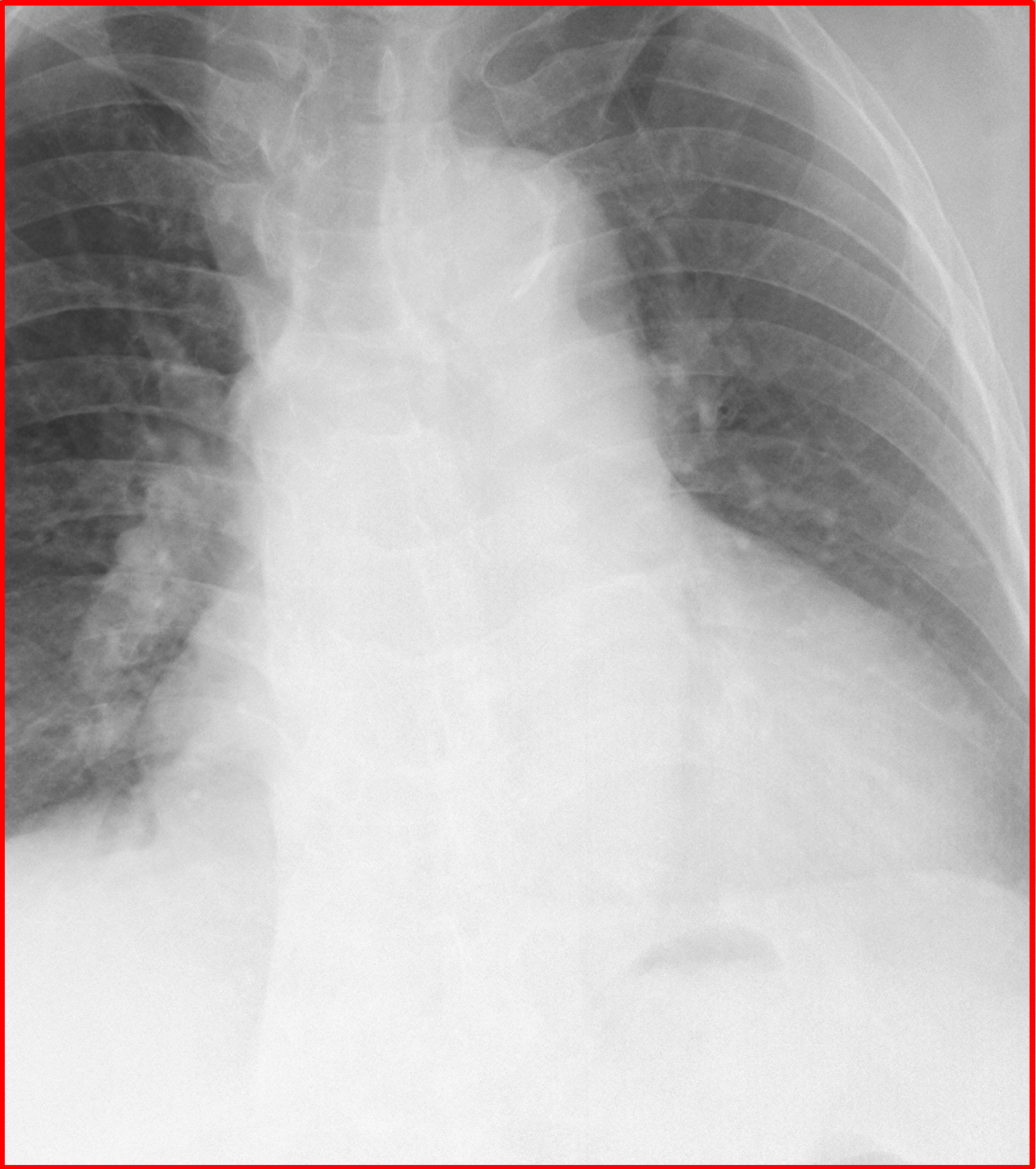}} \hfill %\qquad  %\hfill
    \subfloat[ \label{fig:cardio-example}]{\includegraphics[height=2.5cm]{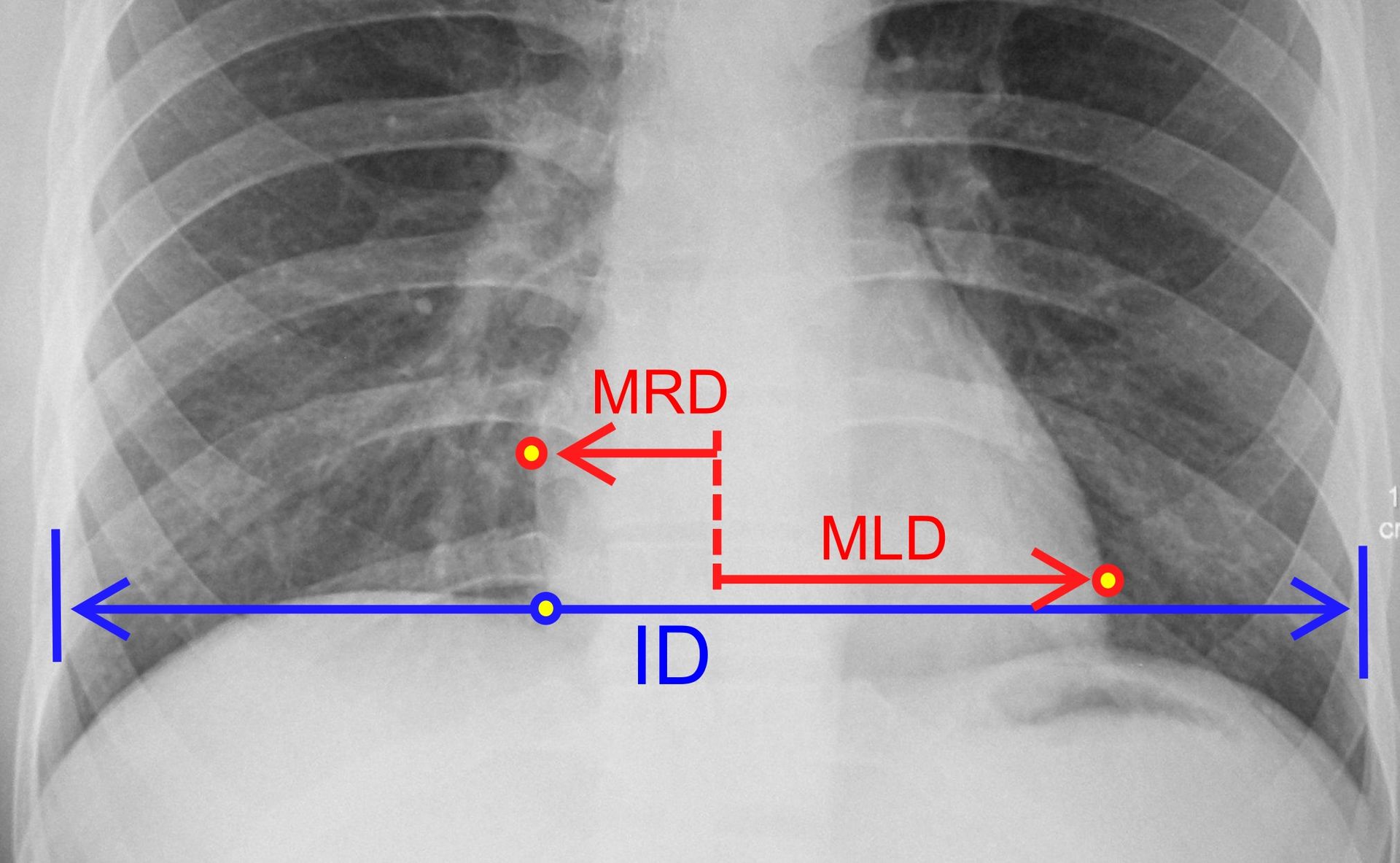}} \hfill
    \subfloat[ \label{fig:ABCDE-cardio}]{\includegraphics[height=2.5cm]{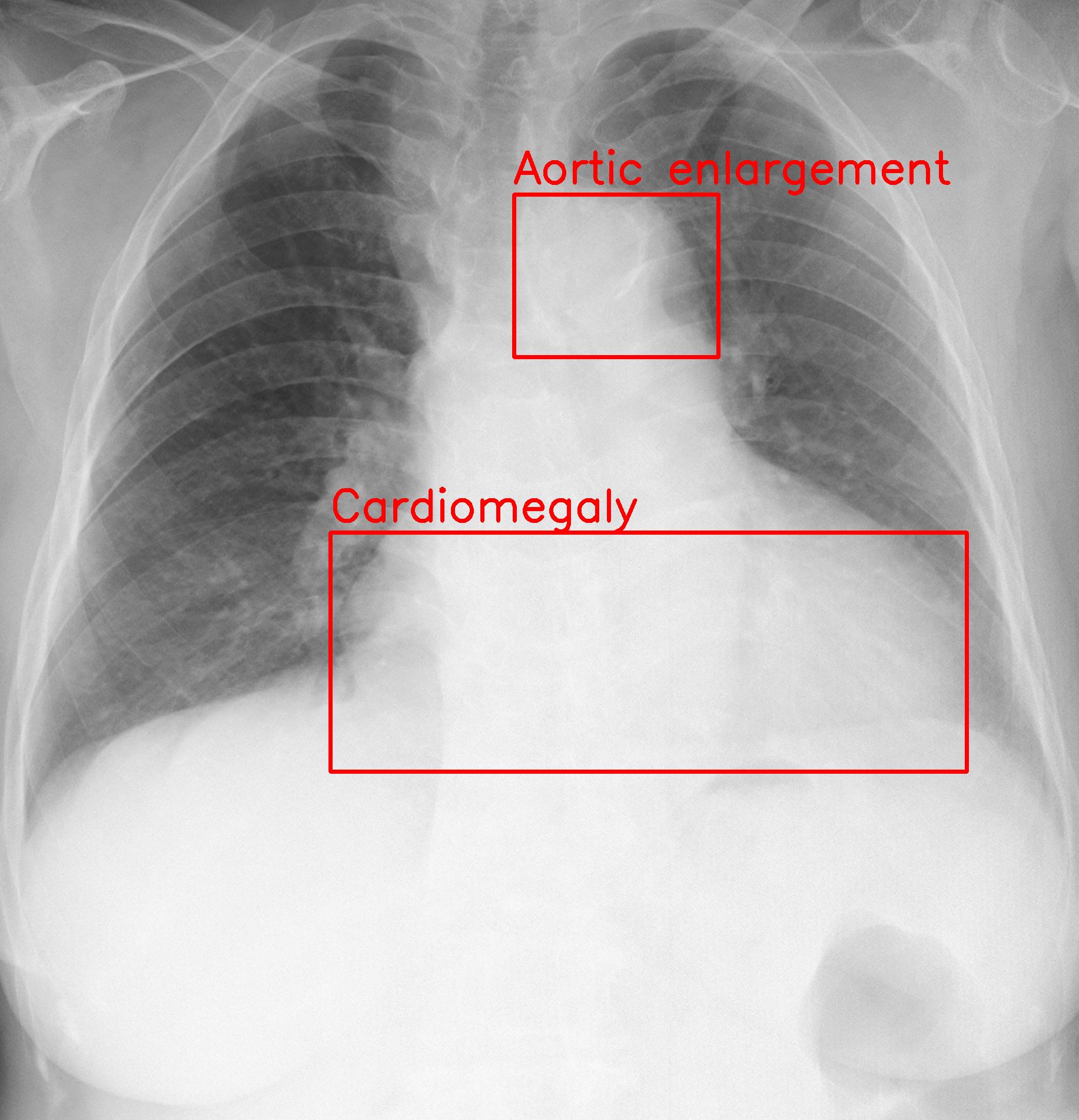}} \hfill
\end{center}
   \caption{ABCDE process ensures that doctors analyze each region in detail \cite{thim2012initial}. In our user trials, we show an example from the VinDr-CXR dataset with aortic enlargement and cardiomegaly in the circulation region. Initially, the circulation region in yellow in the thumbnail view is expanded and shown to ensure a detailed analysis (System 2) of the affected areas. Both aortic enlargement and cardiomegaly require additional context for proper assessment (Fig.\ref{fig:cardio} and \ref{fig:zoom-eval}). Upon selecting these diseases, the view expands to include the thoracic region (Fig.\ref{fig:b1-abcde}) as observed in our analysis using deep learning methods while excluding unrelated parts of the image as shown in the figure. (a) Original Image (b) Regions corresponding to ABCDE is able to cover the entire region and brings focus on all parts of thoracic region (c) Circulation Region (d) Circulation region  expanded in ABCDE workflow (e) Certain diseases are better evaluated by observing entire thoracic region. For example. A cardiothoracic ratio $\geq$ 0.50 indicates cardiomegaly, where cardio-thoracic ratio is $\frac{MRD+MLD}{ID}$. Image Source (Wikipedia). %Certain diseases in circulation region require observing thoracic region.
   (f) Anticipating the diseases based on their characteristics can help in human machine interaction. Cardiomegaly requires larger context but other diseases nodules require to be zoomed for easier recognition. Computational model of human visual can help determine additional context for verification. Screenshots are present in supplementary data.
   }
   %\Description{Design with the ABCDE process showing the Airway.}
\label{fig:abcde_process}
\end{figure*}

\iffalse
\begin{figure*}[h]
\begin{center}   
    \fbox{
    \includegraphics[width=0.55\linewidth]{images/sceenshots/ABCDE_Cardiomegaly.png}   
    }
\end{center}
   \caption{In our user trials, we show an example from the VinDr-CXR dataset with aortic enlargement and cardiomegaly in the circulation region. Initially, the circulation region in yellow in the thumbnail view is expanded and shown to ensure a detailed analysis (System 2) of the affected areas. Both aortic enlargement and cardiomegaly require additional context for proper assessment (Fig.\ref{fig:cardio} and \ref{fig:zoom-eval}). Upon selecting these diseases, the view expands to include the thoracic region (Fig.\ref{fig:b1-abcde}) as observed in our analysis using deep learning methods while excluding unrelated parts of the image as shown in the figure.}
   %\Description{Design with the ABCDE process showing the Airway.}
\label{fig:abcde_process}
\end{figure*}
\fi

Clinicians who utilized the standardized ABCDE analysis approach preferred our method, which integrates this framework into the workflow and allows for closer examination of expanded regions. Participants from urban settings conducted more comprehensive ABCDE analyses, whereas those working in rural areas tended to rely on system 1 thinking or symptom-specific approaches. Clinicians who forgo detailed system 2 analyses, particularly in the absence of radiologists’ reports, can be more susceptible to diagnostic errors. The interviews highlight the potential of AI systems in rural healthcare settings, where clinicians often lack access to detailed reports and are more susceptible to errors such as inattentional blindness in busy clinical environments. Clinicians generally considered the ABCDE approach as a better method. The questions in the VQA-RAD dataset indicate the preference to use the system in a way that mirrored their analysis process, as the questions aligned with different steps in their evaluation, including those needed for intermediate assessments but not necessarily for the final report. Additionally, participants highlighted the tool’s effectiveness in identifying diseases when affected regions are presented with enhanced visual detail, emphasizing the significance of the third part of our study on effective presentation.

\subsection{Diagnostic Errors (RQ:2)}

In this section, we examine common characteristics of diseases likely to be missed and their relationship to existing research on human visual limitations \cite{dayanandan2024dual}. While earlier studies indicate prioritization of conditions that are prone to be overlooked, more recent research focusing on clinicians highlights the importance of prioritizing cases that require urgent intervention or are potentially fatal \cite{dayanandan2024enabling}. Additionally, we explore methods for effectively presenting machine-generated diagnoses to clinicians to enhance usability and clinical decision-making.

\textit{\textbf{(1)}} Inattentional blindness as a source for error is evident in many responses including for nodules which account for 43\% of malpractice claims related to chest imaging \cite{gefter2022special}. While earlier studies cite the small size and unpredictable location of affected regions as reasons for missed nodules \cite{dayanandan2024enabling}, our study identifies additional contributing factors \textit{".. generally we don't order x-ray to find nodules.. we don't look at x-ray's that way.. we don't thoroughly read the x-ray's for nodules and all.. if we suspect the patient is having nodules.. we go for a CT scan.." (P11)}, \textit{".. if a person is missing nodules.. it is not because they do not know to look for a nodule.. it is because.. they are not paying attention .. or maybe they do not have enough practice .. (P15)"}. The errors due to inattentive blindness was also mentioned for other diseases \textit{".. no one takes an x-ray for hernia.. " (P10)} and \textit{".. hilum at times you miss.. you see the lungs and miss that hilum is broad.. actually you should not be missing.. (P2)"}.  Clinicians in rural areas may be more vulnerable to these errors, as they often do not receive a report with the X-ray and primarily rely on analysis based on the patient's symptoms \textit{".. most of the times we don't get a report for the x-ray with the x-ray film.. the general consensus is that all the doctors know how to read an x-ray .. (P11)"}. Existing research also indicates that radiologists could recognize 80\% of diagnostic errors when pointed out \cite{gefter2022special}, underscoring inattentional blindness as a significant factor contributing to diagnostic errors.

\begin{figure*}[h]
\begin{center}   
    \subfloat[ \label{fig:pneumothorax-example}]{\includegraphics[height=3.1cm]{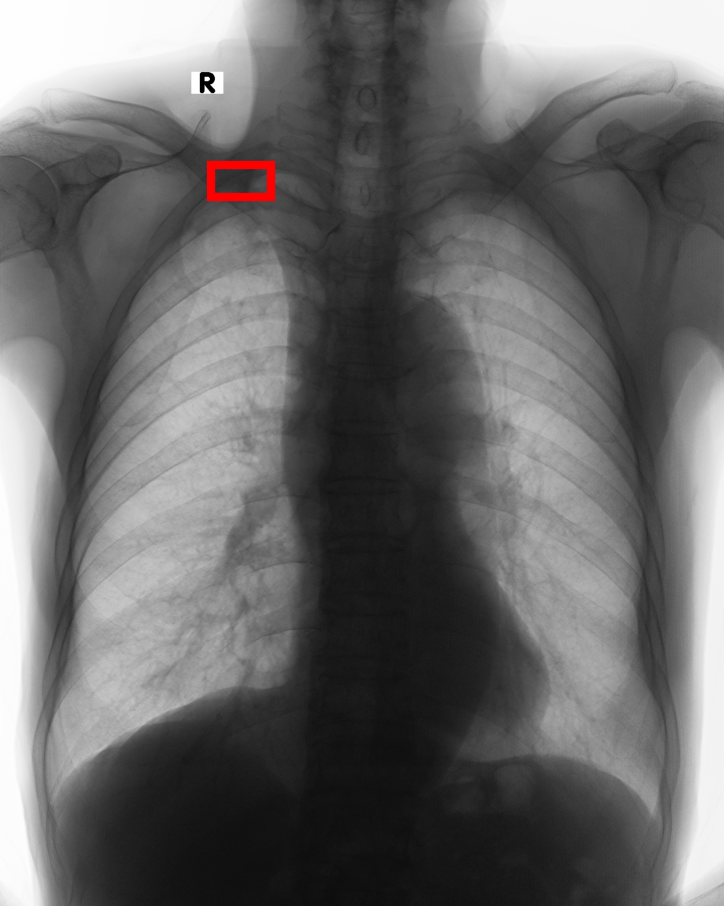}} \hfill
    \subfloat[ \label{fig:pneumothorax-example1}]{\includegraphics[height=3.1cm]{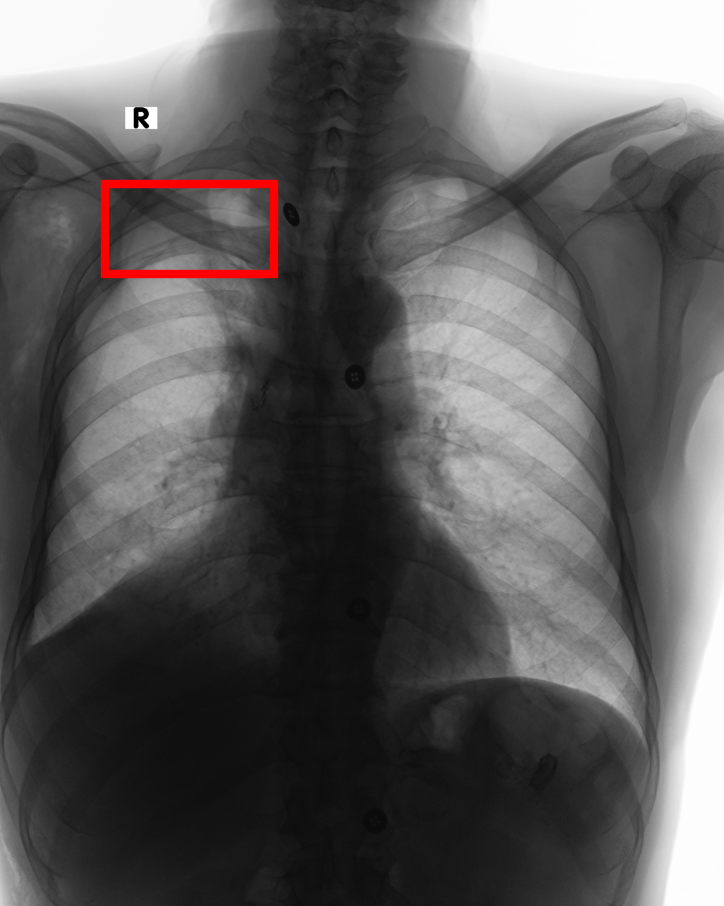}} \hfill
    \subfloat[ \label{fig:pneumothorax-example2}]{\includegraphics[height=3.1cm]{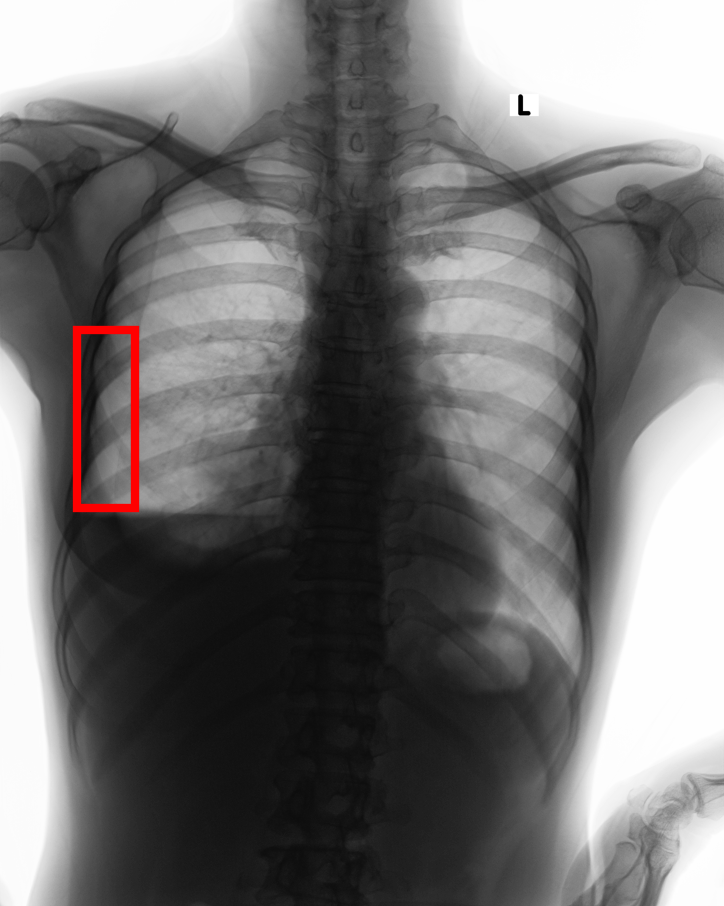}} \hfill
    \subfloat[ \label{fig:pneumothorax-example3}]{\includegraphics[height=3.1cm]{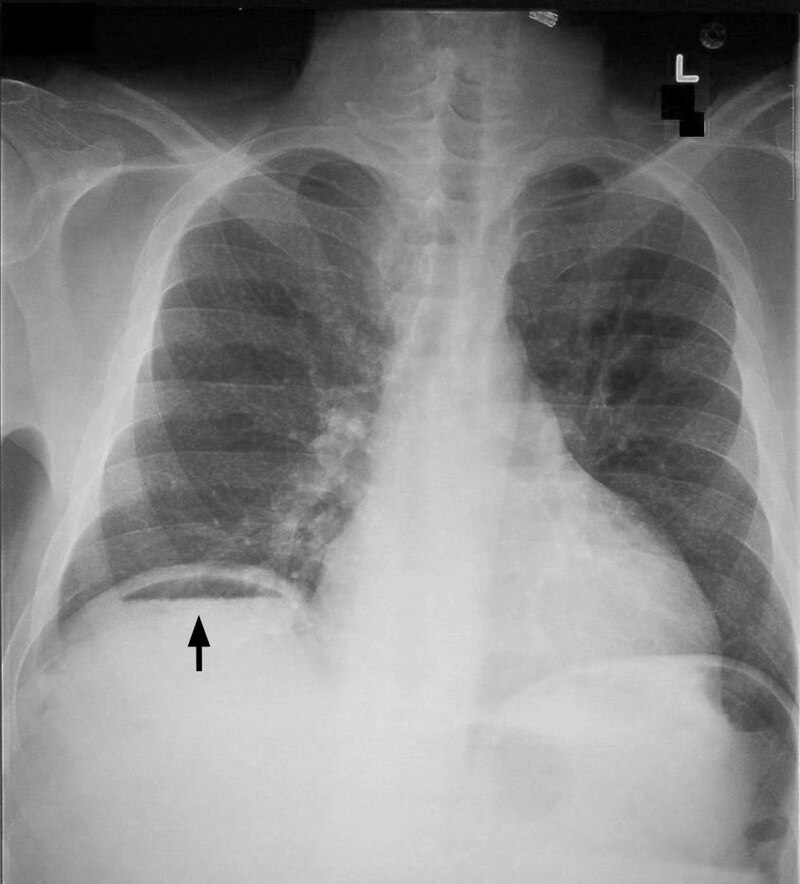}} \hfill
    \subfloat[ \label{fig:cardio-example}]{\includegraphics[height=3.1cm]{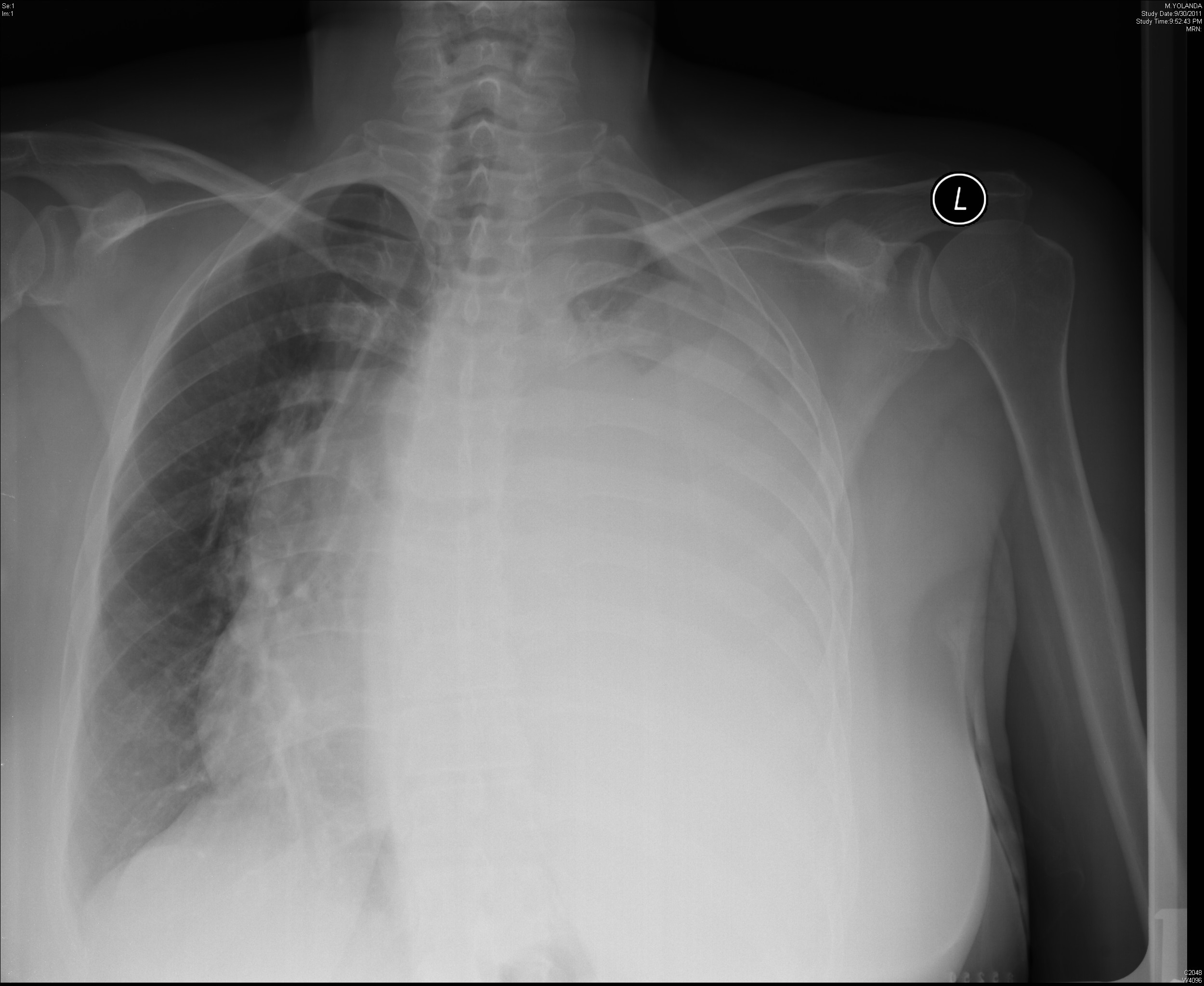}} \hfill
\end{center}
   \caption{Some examples of likely to be missed out diseases (a,b) Pneumothorax in apice from VinDr-CXR dataset (c) Pneumothorax in the border from VinDr-CXR dataset (d) Pneumoperitoneum is abnormal presence of air or other gas in the peritoneal cavity Image Source (Wikipedia).  (d) A massive left pleural effusion displacing the heart and trachea to the right in case of mediastinal (compartment or the thoracic cavity between the pleural sacs of the right and left lungs) shift Image Source (Wikipedia).}
\label{fig:penumothorax}
\end{figure*}

\textit{\textbf{(2)}} Clinicians found it challenging to diagnose many diseases that appear along the lung borders \textit{"..  hiatus hernia at the base.. it may not be seen unless it is too large .. (P4)"},  \textit{"..  rising of the diaphragm.. the right diaphragm is elevated and people don't see it.. hernia.. pneumoperitonium .. or collection of fluid between diagphragm or liver.. (P2)"}. Some diseases occur rarely, which may have contributed to them being overlooked \textit{".. air under diapraghm typically occurs after a surgery.. laproscopic surgery or so .. (P4)"}. We observed that diseases requiring the detection of subtle patterns with low contrast are challenging to diagnose \textit{".. ILD is difficult to diagnose .. they occur like ground glass or tree in bud appearance.. this is common in TB.. .. pleural effusion you can make out.. you will see opacity.. it will be very marked and extensive.. with clear demarcation .. (P6)"}, and \textit{".. shadows might superimpose in some regions.. it wont be clear then.. in the upper zone.. the clavicle also hampers.. (P6) "} and \textit{"..  pneumonia is difficult to diagnose.. the patches are not visible sometimes .. (P10)"}. Diseases that are challenging to diagnose often appear near borders or require detecting subtle textures, which align with the inherent limitations of human vision.

Deep learning models excel in detecting subtle texture differences (Fig. \ref{fig:dual-gen-a}) and are not prone to human biases. Unlike clinicians, who may inadvertently overlook additional conditions after identifying a primary disease, deep learning models typically do not exhibit this vulnerability. Multi-label dependencies have been shown to enhance diagnostic accuracy in AI systems \cite{ge2020improving, sun2023label}, and various techniques have been developed to address the limitations of AI models in diagnosing rare diseases \cite{chen2021multi, holste2022long, wang2022automated, zhang2023knowledge, holste2023cxr, haque2023effect, calisto2020breastscreening}. The thematic analysis of clinicians' descriptions of difficult-to-diagnose diseases aligns with known limitations of human vision. For instance, human vision tends to focus on overall shapes while neglecting subtle textures or fine details at boundaries (Fig. \ref{fig:navon-dataset}, \ref{fig:navon-consistency}). Pneumothorax occurring at the upper corners along the boundary (Fig.\ref{fig:penumothorax}), has been highlighted in prior studies as likely to miss out \cite{gefter2022special}; however, such conditions can be easily identified with appropriate magnification.  
We also note a correlation between diseases that are often missed—such as pneumothorax, which requires urgent attention, or nodules indicative of early-stage lung cancer, which can be fatal—and their frequent involvement in legal malpractice cases \cite{gefter2022special}. In contrast, mild pleural effusion, typically observed at the boundary as blunting of the costophrenic angle, is also prone to be overlooked but does not require intervention \cite{dayanandan2024enabling} is rarely reported in such legal cases \cite{gefter2022special}. AI systems have the potential to complement clinicians, acting as a "second pair of eyes" to ensure more accurate and comprehensive diagnoses.

\subsection{Diagnosis Evaluation (RQ:3)}

In this section, we show that the contextual information required for evaluating machine diagnoses varies based on the characteristics of diseases, and later, we show that the computational model of human vision can capture many of these details. In the first part of this section, we describe the computational approach to identify the context around affected regions for diagnosing various diseases, focusing on the effective receptive field of the deep learning model and the resolution at which they achieve optimal performance. In the final part, we present participants' feedback on the contextual information around the affected areas that clinicians need for accurate diagnosis.

\subsubsection{Computational Model of Human Vision} 

The receptive field in human vision denotes the specific region within the sensory space that elicits a neuronal response when stimulated. Similarly, in deep learning, the receptive field (RF) represents the portion of an input image that contributes to the computation of a particular feature, with regions outside the effective receptive field exerting no influence on model predictions \cite{haque2023effect}. For instance, cardiomegaly, a condition characterized by heart enlargement, is diagnosed by assessing the ratio of the heart to the overall lung region (Fig.\ref{fig:cardio}). Clinicians must observe the heart in relation to the entire lung field, necessitating a complete view of the thoracic region rather than the heart alone. Similarly, in deep learning models, accurate predictions necessitate the entire thoracic area within the effective receptive field for accurate prediction, which is achievable at lower image resolutions and has been shown in many studies \cite{haque2023effect,sabottke2020effect} (Appendix contains Table \ref{tab:model-performance-dense} and Table \ref{tab:model-performance-efb4} from \cite{haque2023effect} on MIMIC-CXR dataset \cite{johnson2019mimic} and Table \ref{tab:model-performance-prior-resnet} from \cite{sabottke2020effect} on the Chest X-Ray8 dataset \cite{wang2017chestx}). Diseases necessitating a larger context for accurate diagnosis can be identified by correlating model performance across varying resolutions, eliminating the need for detailed disease-specific insights. We trained models separately for each disease (Table \ref{tab:model-performance-effnet-vindr}) to better capture their characteristics along with early stopping to reduce computation, in contrast to existing studies that train models on multiple diseases together. Our study also focuses on diseases that were not part of earlier studies. For aortic enlargement, though the affected region is comparatively small, deep learning models performed better with smaller image sizes compared to calcification, which performed better on a higher-resolution image. 

\begin{table}[h]
\caption{Area Under Curve (AUC) scores for EfficientNet-B4 trained on down scaled images on VinDr-CXR dataset. We train the models separately for different diseases to capture disease characteristics better, whereas existing studies and baseline code train models for overall accuracy. We use the baseline code from third place solution from the competition which use 1024x1024 resolution and use a patience of one for early stopping to reduce computation.}
\label{tab:model-performance-effnet-vindr}
\vskip 0.12in
\centering
\begin{tabular}{llll}
\toprule
Finding & 256x256 & 512x512 & 1024x1024 \\
\midrule
Aortic enlargement & \textbf{0.87852} & 0.84916 & 0.80065 \\
Atelectasis & 0.85537 & \textbf{0.86459} & 0.72605 \\
Calcification & 0.84789 & \textbf{0.85721} & 0.82633 \\
Cardiomegaly & \textbf{0.91834} & 0.91108 & 0.86363 \\
Clavicle fracture & 0.7068 & \textbf{0.82155} & 0.6074 \\
Consolidation & 0.88229 & \textbf{0.91126} & 0.70836 \\
Emphysema & 0.97242 & \textbf{0.98354} & 0.93271 \\
Enlarged PA & \textbf{0.83727} & 0.81722 & 0.76512  \\
ILD & \textbf{0.86659} & 0.86396 & 0.77125 \\
Infiltration & 0.89115 & \textbf{0.918} & 0.78902 \\
Lung Opacity & \textbf{0.83696} & 0.82819 & 0.77999 \\
Lung cavity & 0.84754 & \textbf{0.88473} & 0.78837 \\
Lung cyst & 0.95147 & 0.87658 & \textbf{0.97215} \\
Mediastinal shift & \textbf{0.92827} & 0.92747 & 0.74507 \\
Nodule/Mass & 0.79119 & \textbf{0.8179} & 0.70267 \\
Pleural effusion & \textbf{0.95108} & 0.93354 & 0.75953 \\
Pleural thickening & \textbf{0.87404} & 0.87125 & 0.80461 \\
Pneumothorax & \textbf{0.91534} & 0.90992 & 0.71637 \\
Pulmonary fibrosis & 0.84525 & \textbf{0.84785} & 0.74104 \\
Rib fracture & 0.83324 & \textbf{0.90292} & 0.78886 \\
Other lesion & \textbf{0.84723} & 0.82787 & 0.79439 \\
COPD & 0.9428 & 0.92195 & \textbf{0.99466} \\
Lung tumor & 0.79531 & \textbf{0.8179} & 0.66313 \\
Pneumonia & 0.89147 & \textbf{0.90468} & 0.7558 \\
Tuberculosis & 0.88065 & \textbf{0.91263} & 0.75852 \\
Overall AUC & 0.84159 & \textbf{0.84824} & 0.75706 \\
\bottomrule
\end{tabular}

\end{table}

\subsubsection{Clinician's Feedback} 

\begin{figure*}[h]
\begin{center}   
    \subfloat[ \label{fig:aortic-OnlySize}]{\includegraphics[height=5.0cm]{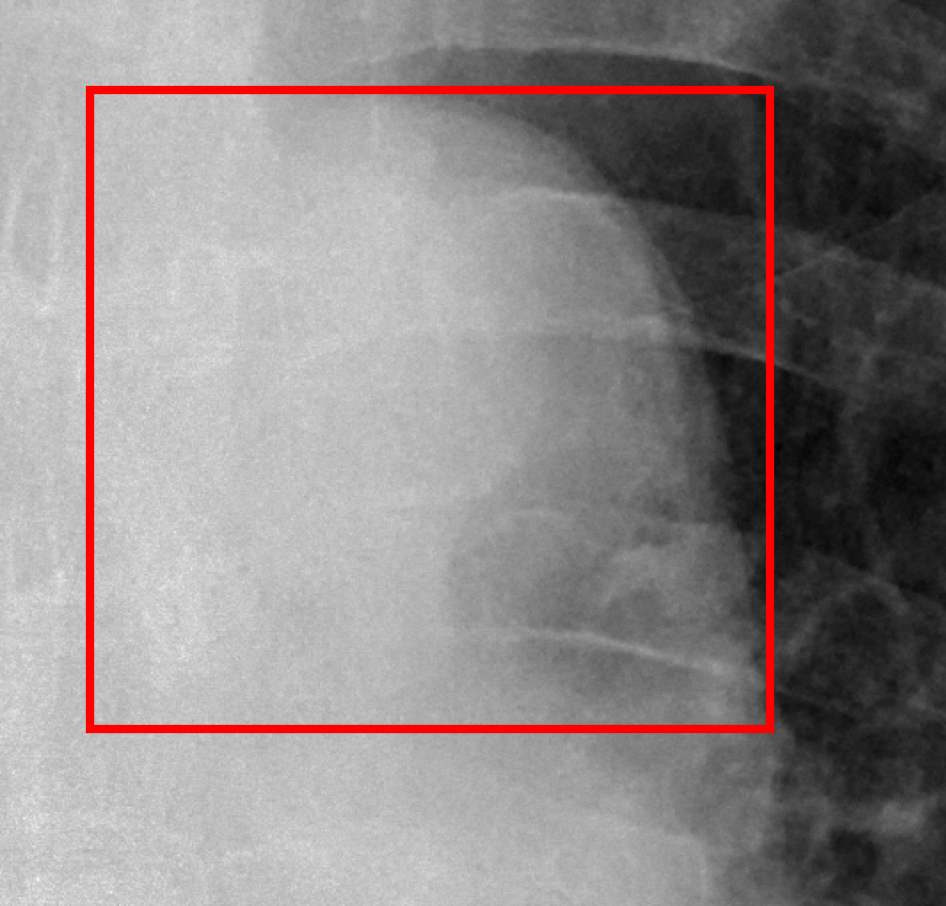}} \hfill
    \subfloat[ \label{fig:aortic-PartialSize}]{\includegraphics[height=5.0cm]{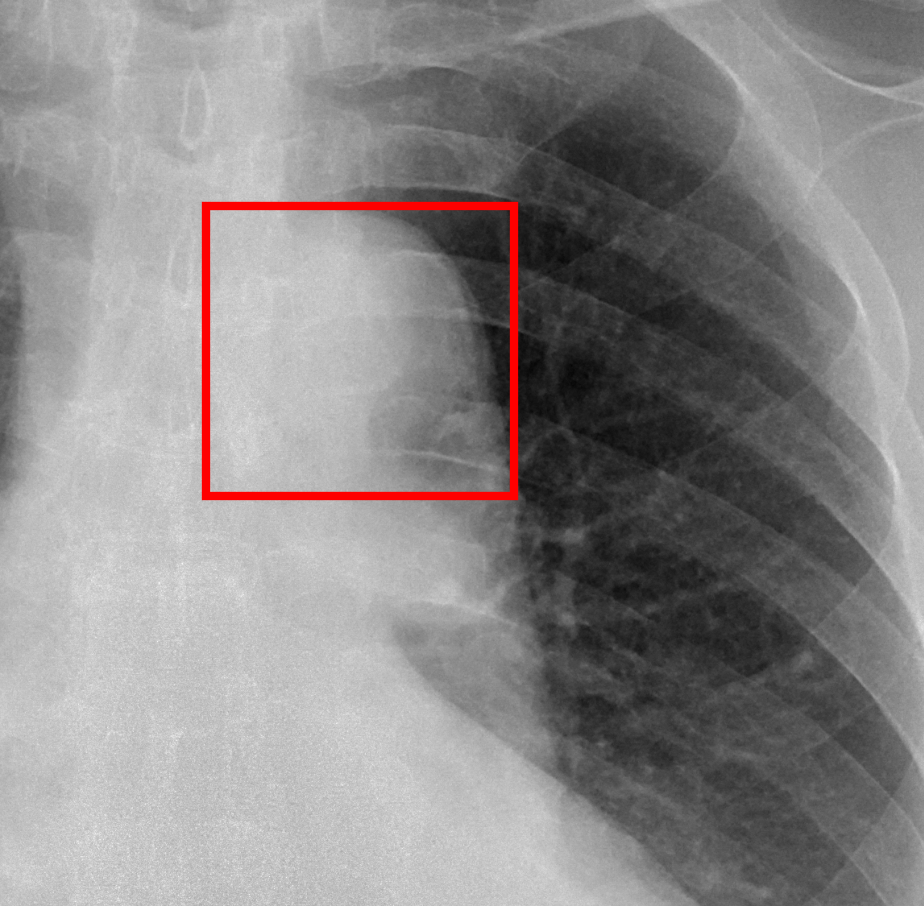}} \hfill
    \subfloat[ \label{fig:aortic-FullSize}]{\includegraphics[height=5.0cm]{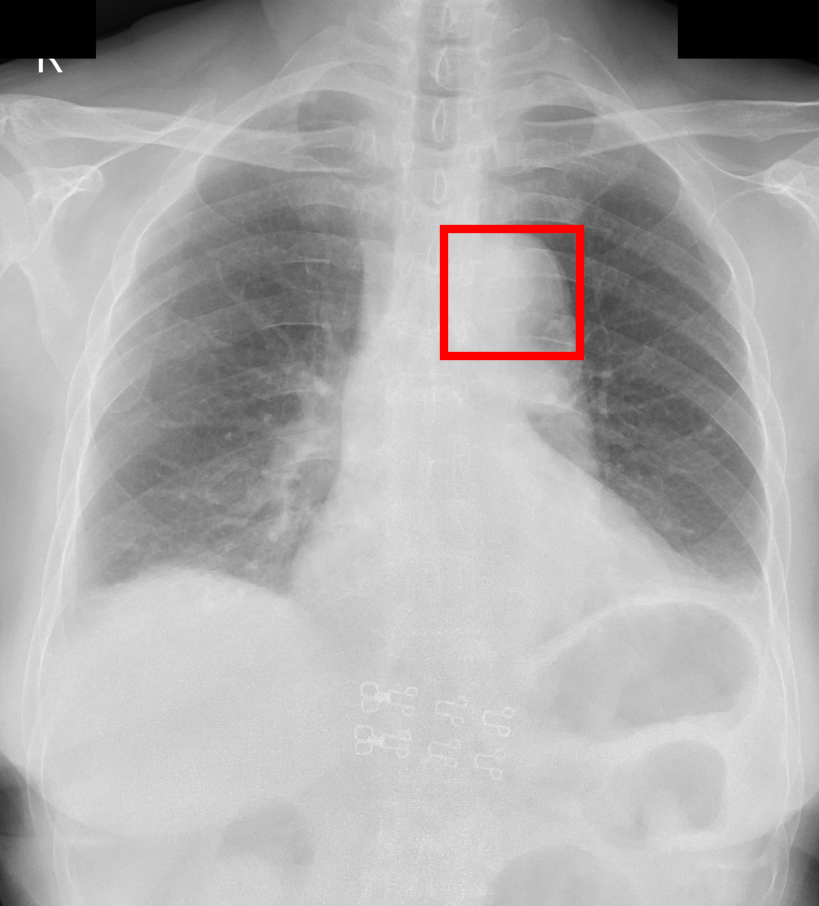}} \hfill

    \subfloat[ \label{fig:Calcification-OnlySize}]{\includegraphics[height=4.2cm]{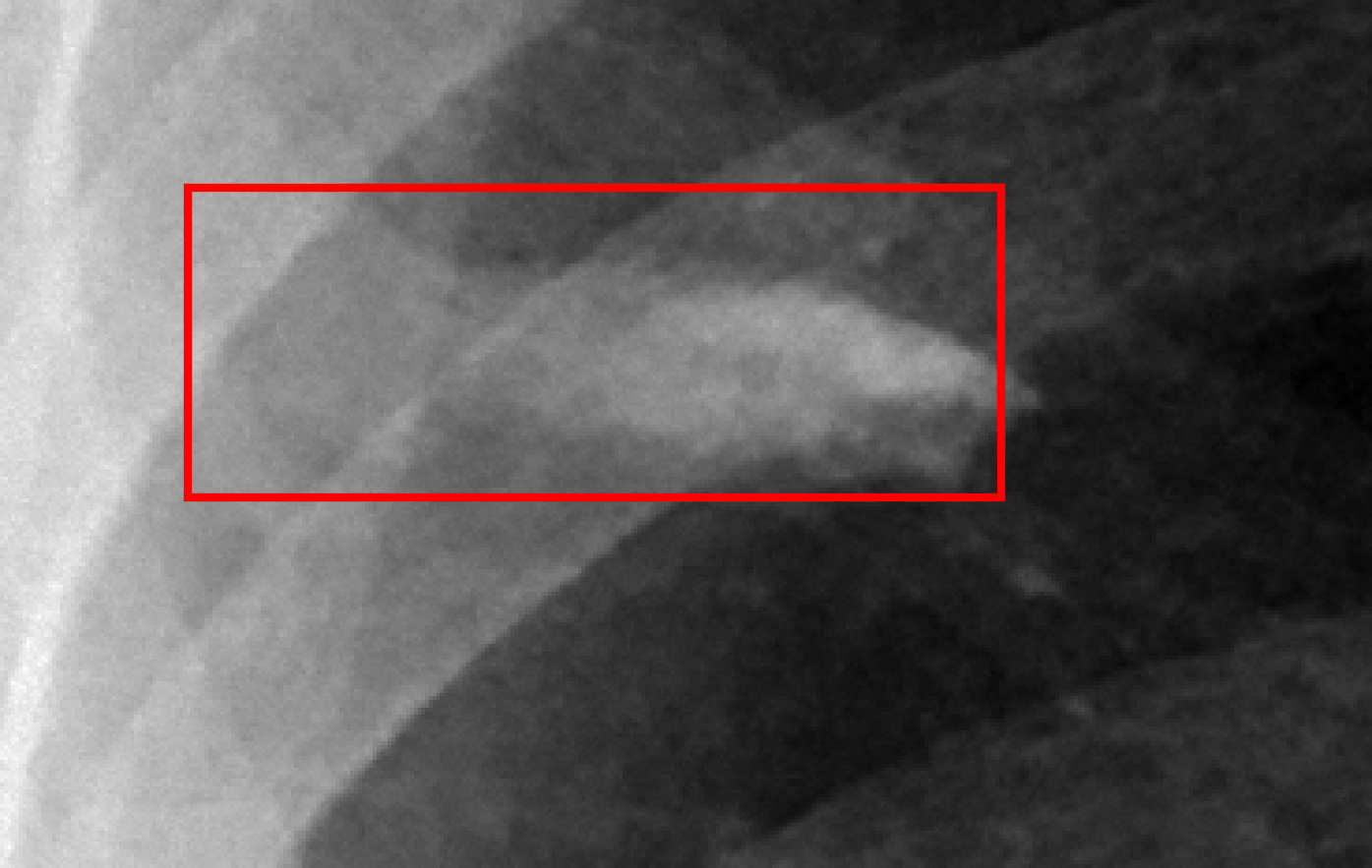}} \hfill
    \subfloat[ \label{fig:Calcification-PartialSize}]{\includegraphics[height=4.2cm]{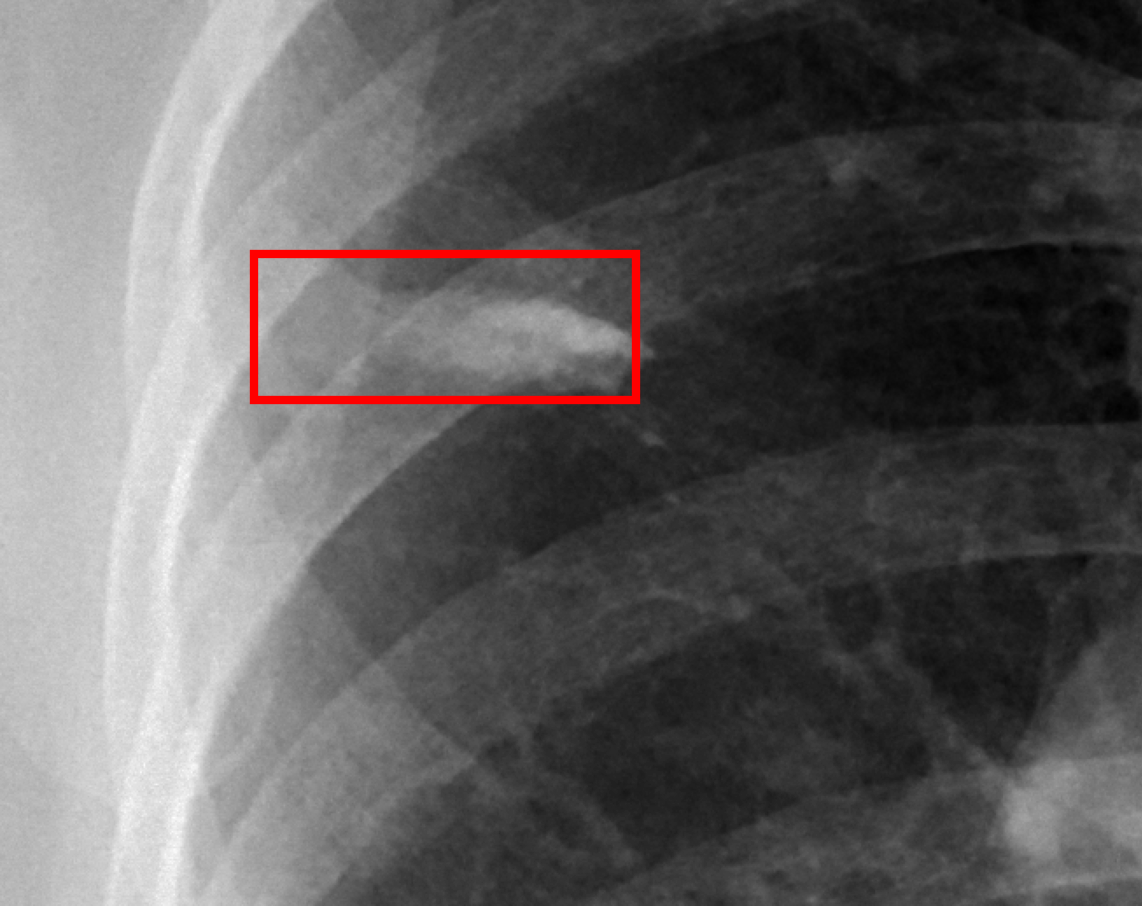}} \hfill
    \subfloat[ \label{fig:Calcification-FullSize}]{\includegraphics[height=4.2cm]{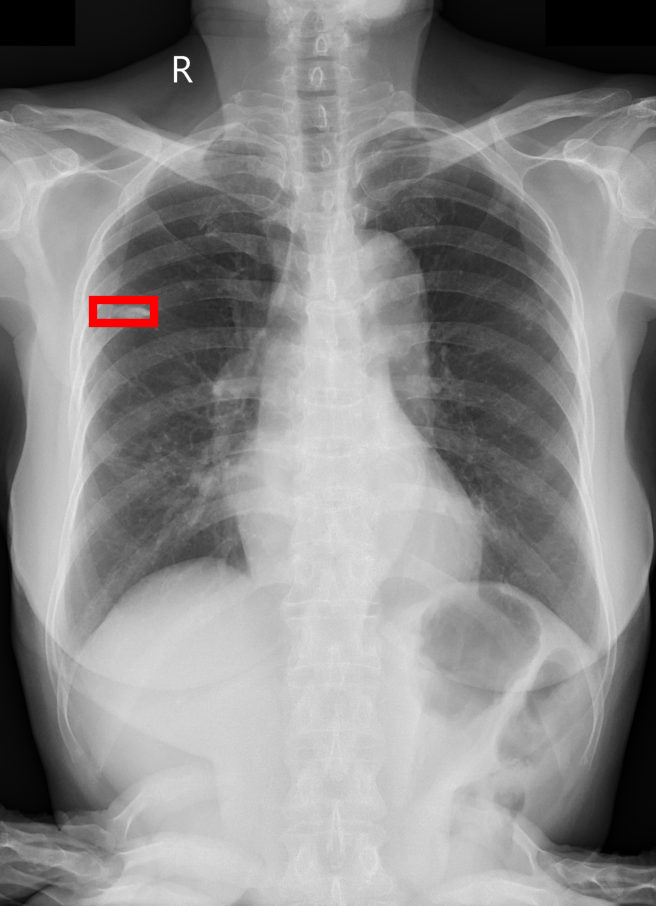}} \hfill
\end{center}
   \caption{ First row shows Aortic Enlargement and second row shows Calcification in different zoom levels (a) Affected region only shown for Aortic Enlargement (b) Affected region with some context for Aortic Enlargement (c) Affected region shown for Aortic Enlargement on full size x-ray (d) Affected region only shown for Calcification (e) Affected region with some context for Calcification (f) Affected region shown for Calcification on full size x-ray}
   %\Description{Evaluation of Aortic Enlargement and Calcification.}
\label{fig:zoom-eval}
\end{figure*}

We presented participants with three viewing options: (a) a zoomed-in view of the exact affected region, (b) additional context surrounding the affected region, and (c) the affected region marked on the original image as shown in Fig.\ref{fig:zoom-eval}. Clinicians preferred a smaller context for calcification and a larger context or a full-image view for aortic enlargement, which involves size comparisons \textit{“..the last one is alright in aortic enlargement.. for calcification.. it is clearer in the first.. in the last one it is too small.." (P2)} and \textit{"aortic enlargement.. show it in full view.. calcification full view is difficult.."(P12)}. This demonstrates that different diseases require varying contextual representations for accurate assessment. Further, we asked clinicians which diseases they believed required zoomed-in views versus larger context. Their responses generally aligned with our findings based on deep learning models as computational representations of human vision \textit{“.. fracture.. can be seen better with zooming.. cardiomegaly on full x-ray.. as cardiomegaly is a ratio.. (P2)”} and \textit{"fractures.. a zoomed in picture makes more of a sense because it is not clear.. when you zoom in.. there can be a false kind of interpretation also.. the small things can appear big.. you need the one on top.. we always see it from the overall x-ray.. (P12)}. The participant’s answers reflected our approach to providing sufficient context for showing diagnosis regions; for example, the diagnosis of cardiomegaly is based on the ratio of heart area to chest area, referred to as the cardio-thoracic ratio. This requires showing the complete thoracic region, including the location of the heart \textit{“.. the complete thoracic region has to be shown for cardiomegaly .. (P4)”}. One of the participants asked whether a scale can be shown for cardiomegaly to make it easier for clinicians  \textit{“..  can you provide scale to show severity of cardiomegaly ..” (P4)”}. This suggestion shows clinicians to favor the presentation of machine diagnosis using methods that clinicians follow in their practice for easier supervision.

\section{Design Implications}

In this sub-section, we summarize our findings from the qualitative study of the proposed approach from a human-computer interaction perspective.

\textbf{(a) Clinicians prefer systems that mimic their approach of analysis for easy supervision.} Doctors analyze chest X-rays sequentially to ensure that reports accurately capture and describe anomalies, along with any related normal conditions. The participating clinicians found our approach, which replicates their analysis methodology, to be helpful. Our findings align with a recent shift toward explaining decisions in terms of the clinician's analytical process rather than solely focusing on increasing clinicians' trust by explaining AI's decision-making process, as observed in earlier studies on improving AI explainability. This shift may be influenced by clinicians' growing awareness of AI's improved performance in recent years and their preference for adhering to their established diagnostic methods.

\textbf{(b) Clinical systems should focus on difficult to diagnose diseases and those prone to being overlooked.} Our study found that clinicians in busy settings, particularly in rural areas or emergency care, are more likely to rely on an overall assessment or diagnosis based primarily on patient symptoms. This, combined with the absence of medical reports, can result in missed diagnoses that require a more detailed examination of different regions, making clinicians more susceptible to System 2 errors. Our analysis of difficult-to-diagnose diseases revealed that they often align with the inherent limitations of human vision. Clinical support systems should prioritize these conditions to make the systems more complementary for clinicians, and such systems can improve accuracy by prioritizing diseases based on their characteristics and the specific regions where they are most likely to be missed.

\textbf{(c) Clinical systems should provide relevant context for easier assessment and verification of diagnosis.} Diseases possess distinct characteristics, and doctors rely on varied criteria for diagnosis. An intuitive interface should anticipate the contextual information needed for diagnosis and present it appropriately to facilitate efficient and accurate supervision. Deep learning-based computational models of human vision can help identify diseases that require a broader context for accurate detection and interpretation. Identifying and presenting the necessary context enables clinicians to review machine-generated diagnoses more quickly and efficiently, minimizing the need for manual image adjustments like zooming and repositioning to bring the relevant regions into view.

\subsection*{Limitations}
Our approach, which uses deep learning models, captures the overall characteristics of diseases as represented in the dataset. However, these characteristics can vary depending on the source of the dataset. Disease characteristics may differ in specialized or referral hospitals primarily treating severe cases.

\section{Conclusion}

In this paper, we explore methods to enhance clinicians' efficiency and accuracy through the support of clinical imaging systems. Our approach replicates clinicians' analysis procedures, allowing the system to assist them more effectively. This collaborative effort can improve efficiency and accuracy, which are critical in safety-sensitive clinical settings. We also examined patterns in difficult-to-diagnose diseases to prevent errors caused by human biases and limitations. We observe that the diseases located at the lung borders and in central regions with dense, bony structures are common areas of misdiagnosis. Deep learning models can detect subtle patterns better and are not subject to the biases found in human analysis, making them complementary to clinicians. Improvement in clinician efficiency can reduce turnaround times, and improvements in accuracy can result in better patient care.

\clearpage

\section{Appendix}

\begin{table}[h]
\caption{Area Under Curve (AUC) scores for Densenet-121 trained on down scaled images after ImageNet pre-training, in percent ± one standard deviation reproduced from \cite{haque2023effect}. }
\label{tab:model-performance-dense}
\vskip 0.12in
\begin{tabularx}{0.75\linewidth}{lXXXXX}
\toprule
         Finding & 256 × 256 &    512 × 512 &   1024 × 1024 &     2048 × 2048 \\
\midrule
Atelectasis & 80.8 $\pm$ 0.6 & 81.7 $\pm$ 0.2 & \textbf{81.8 $\pm$ 0.4} & 80.9 $\pm$ 0.4 \\
Cardiomegaly & \textbf{81.6 $\pm$ 0.4} & 81.5 $\pm$ 0.5 & 81.2 $\pm$ 0.5 & 79.9 $\pm$ 0.4 \\
Consolidation & 82.1 $\pm$ 0.7 & \textbf{82.7 $\pm$ 0.2} & 82.5 $\pm$ 0.3 & 81.0 $\pm$ 0.2 \\
Edema & 88.9 $\pm$ 0.3 & 89.8 $\pm$ 0.4 & \textbf{90.0 $\pm$ 0.2} & 89.3 $\pm$ 0.3 \\
Enlarged cardiomediastinum & \textbf{73.9 $\pm$ 0.9} & 73.8 $\pm$ 0.8 & 73.3 $\pm$ 0.8 & 72.3 $\pm$ 1.1 \\
Fracture & 66.7 $\pm$ 1.9 & \textbf{67.4 $\pm$ 1.4} & 67.3 $\pm$ 1.3 & 65.6 $\pm$ 0.9 \\
Lung lesion & 73.8 $\pm$ 0.9 & \textbf{75.0 $\pm$ 0.9} & 74.5 $\pm$ 0.7 & 73.8 $\pm$ 0.5 \\
Lung opacity & 74.9 $\pm$ 0.4 & 76.1 $\pm$ 0.2 & \textbf{76.3 $\pm$ 0.3} & 75.4 $\pm$ 0.4 \\
No finding & 85.3 $\pm$ 0.4 & 85.8 $\pm$ 0.4 & \textbf{85.9 $\pm$ 0.3} & 85.3 $\pm$ 0.2 \\
Pleural effusion & 91.9 $\pm$ 0.4 & 92.3 $\pm$ 0.4 & \textbf{92.3 $\pm$ 0.3} & 91.5 $\pm$ 0.4 \\
Pleural other & 80.6 $\pm$ 0.3 & \textbf{82.5 $\pm$ 1.1} & 82.2 $\pm$ 0.7 & 81.6 $\pm$ 0.5 \\
Pneumonia & 71.4 $\pm$ 0.6 & 72.9 $\pm$ 0.4 & \textbf{73.1 $\pm$ 0.5} & 71.6 $\pm$ 0.6 \\
Pneumothorax & 85.6 $\pm$ 1.1 & 87.8 $\pm$ 0.7 & 88.4 $\pm$ 0.9 & \textbf{88.7 $\pm$ 0.4} \\
Support devices & 89.8 $\pm$ 0.4 & 91.8 $\pm$ 0.1 & \textbf{92.3 $\pm$ 0.3} & 92.1 $\pm$ 0.2 \\
Average & 80.5 $\pm$ 0.3 & 81.5 $\pm$ 0.3 & 81.5 $\pm$ 0.3 & \textbf{80.7 $\pm$ 0.1} \\
\bottomrule
\end{tabularx}
\end{table}

\begin{table}[h]
\caption{Area Under Curve (AUC) scores for EfficientNet-B4 trained on down scaled images after ImageNet pre-training, in percent ± one standard deviation reproduced from \cite{haque2023effect}. }
\label{tab:model-performance-efb4}
\vskip 0.12in
\begin{tabularx}{0.75\linewidth}{llllll}
\toprule
         Finding & 256 × 256 &    512 × 512 &   1024 × 1024 &     2048 × 2048 \\
\midrule
Atelectasis     & 81.9 $\pm$ 0.3 & 82.8 $\pm$ 0.4 & \textbf{82.9 $\pm$ 0.3} & 82.5 $\pm$ 0.5 \\
Cardiomegaly    & \textbf{82.4 $\pm$ 0.5} & 82.3 $\pm$ 0.3 & 82.2 $\pm$ 0.4 & 81.6 $\pm$ 0.4 \\
Consolidation   & 82.6 $\pm$ 0.3 & 83.6 $\pm$ 0.3 & \textbf{83.8 $\pm$ 0.2} & 83.3 $\pm$ 0.3 \\
Edema           & 89.7 $\pm$ 0.3 & 90.5 $\pm$ 0.2 & \textbf{90.6 $\pm$ 0.2} & 90.4 $\pm$ 0.4 \\
Enlarged cardiomediastinum & 73.8 $\pm$ 0.9 & \textbf{74.1 $\pm$ 1.0} & 74.0 $\pm$ 0.9 & 73.3 $\pm$ 0.8 \\
Fracture        & 67.0 $\pm$ 1.4 & 69.5 $\pm$ 1.2 & \textbf{70.8 $\pm$ 1.7} & 69.6 $\pm$ 1.8 \\
Lung lesion     & 74.9 $\pm$ 1.3 & 76.6 $\pm$ 0.8 & 77.7 $\pm$ 0.6 & \textbf{78.2 $\pm$ 0.6} \\
Lung opacity    & 76.3 $\pm$ 0.2 & 77.4 $\pm$ 0.2 & \textbf{77.8 $\pm$ 0.1} & 77.3 $\pm$ 0.5 \\
No finding      & 86.1 $\pm$ 0.2 & 86.7 $\pm$ 0.2 & \textbf{86.8 $\pm$ 0.2} & 86.5 $\pm$ 0.2 \\
Pleural effusion& 92.5 $\pm$ 0.3 & 92.9 $\pm$ 0.2 & \textbf{93.0 $\pm$ 0.2} & 92.6 $\pm$ 0.2 \\
Pleural other   & 81.9 $\pm$ 0.1 & 84.1 $\pm$ 0.8 & \textbf{84.7 $\pm$ 0.7} & 84.1 $\pm$ 0.6 \\
Pneumonia       & 73.2 $\pm$ 0.4 & 74.8 $\pm$ 0.4 & \textbf{75.3 $\pm$ 0.4} & 75.0 $\pm$ 0.5 \\
Pneumothorax    & 88.0 $\pm$ 0.4 & 90.6 $\pm$ 0.2 & 91.9 $\pm$ 0.3 & \textbf{92.0 $\pm$ 0.3} \\
Support devices & 91.4 $\pm$ 0.1 & 93.2 $\pm$ 0.2 & \textbf{93.7 $\pm$ 0.2} & 93.7 $\pm$ 0.2 \\
Average         & 81.5 $\pm$ 0.1 & 82.8 $\pm$ 0.1 & \textbf{83.2 $\pm$ 0.1} & 82.9 $\pm$ 0.1 \\
\bottomrule
\end{tabularx}
\end{table}

\begin{table}[h]
\caption{Area Under Curve (AUC) scores for ResNet-34 reproduced from \cite{sabottke2020effect} on Chest X-Ray8 dataset \cite{wang2017chestx}.}
\label{tab:model-performance-prior-resnet}
\vskip 0.12in
\begin{tabular}{llllll}
\toprule
Finding & 256 × 256 & 320 × 320 & 448 × 448 & 512 × 512 & 600 × 600 \\
\midrule
Emphysema & 0.916 & 0.935 & 0.931 & \textbf{0.936} & 0.933 \\
Cardiomegaly & 0.916 & \textbf{0.927} & 0.922 & 0.894 & 0.882 \\
Hernia & 0.804 & \textbf{0.838} & 0.812 & 0.687 & 0.75 \\
Atelectasis & 0.882 & 0.887 & \textbf{0.893} & 0.87 & 0.853 \\
Edema & 0.917 & \textbf{0.924} & 0.916 & 0.905 & 0.909 \\
Effusion & 0.913 & 0.913 & \textbf{0.919} & 0.902 & 0.901 \\
Mass & 0.879 & 0.886 & \textbf{0.894} & 0.862 & 0.847 \\
Nodule & 0.827 & 0.854 & \textbf{0.868} & 0.836 & 0.833 \\
\bottomrule
\end{tabular}

\end{table}

\clearpage

\bibliographystyle{ACM-Reference-Format}
\bibliography{sample-base}

\end{document}